\documentclass[11pt]{article}
\usepackage{graphicx}
\usepackage{amssymb}
\usepackage{amscd}
\usepackage{mathrsfs}
\usepackage{longtable,lscape}
\usepackage{amsthm}
\usepackage{amsfonts}
\usepackage{amsmath}
\usepackage{bbm}
\usepackage{float}
\usepackage{url}

\newcommand{\captionfonts}{\footnotesize}
\makeatletter  % Allow the use of @ in command names
\long\def\@makecaption#1#2{%
  \vskip\abovecaptionskip
  \sbox\@tempboxa{{\captionfonts #1: #2}}%
  \ifdim \wd\@tempboxa >\hsize
    {\captionfonts #1: #2\par}
  \else
    \hbox to\hsize{\hfil\box\@tempboxa\hfil}%
  \fi
  \vskip\belowcaptionskip}
\makeatother 
\begin{document}
\title{Spin and Wind Directions II: A Bell State Quantum Model}
\author{Diederik Aerts$^1$, Jonito Aerts Argu\"elles$^2$, Lester Beltran$^3$, Suzette Geriente$^4$, \\ Massimiliano Sassoli de Bianchi$^{1}$, Sandro Sozzo$^{5}$  and Tomas Veloz$^1$ \vspace{0.5 cm} \\ 
        \normalsize\itshape
        $^1$ Center Leo Apostel for Interdisciplinary Studies, 
         Brussels Free University \\ 
        \normalsize\itshape
         Krijgskundestraat 33, 1160 Brussels, Belgium \\
        \normalsize
        E-Mails: \url{diraerts@vub.ac.be,msassoli@vub.ac.be},\\\url{tveloz@gmail.com}
          \vspace{0.5 cm} \\ 
        \normalsize\itshape
        $^2$ KASK and Conservatory, \\
        \normalsize\itshape
         Jozef Kluyskensstraat 2, 9000 Ghent, Belgium
        \\
        \normalsize
        E-Mail: \url{jonitoarguelles@gmail.com}
	  \vspace{0.5 cm} \\ 
        \normalsize\itshape
        $^3$ 825-C Tayuman Street, \\
         \normalsize\itshape
        Tondo, Manila, The Philippines
         \\
        \normalsize
        E-Mail: \url{lestercc21@gmail.com}
	  \vspace{0.5 cm} \\ 
        \normalsize\itshape
        $^4$ Block 28 Lot 29 Phase III F1, \\ 
         \normalsize\itshape
        Kaunlaran Village, Caloocan City, The Philippines
         \\
        \normalsize
        E-Mail: \url{sgeriente83@yahoo.com}
	  \vspace{0.5 cm} \\ 
        \normalsize\itshape
        $^5$ School of Management and IQSCS, University of Leicester \\ 
        \normalsize\itshape
         University Road, LE1 7RH Leicester, United Kingdom \\
        \normalsize
        E-Mail: \url{ss831@le.ac.uk} 
       	\\
              }
\date{}
\maketitle
\begin{abstract}
\noindent
In the first half of this two-part article \cite{ass2017}, we analyzed a cognitive psychology experiment where participants were asked to select pairs of directions that they considered to be the best example of {\it Two Different Wind Directions}, and showed that the data violate the CHSH version of Bell's inequality, with same magnitude as in typical Bell-test experiments in physics. In this second part, we complete our analysis by presenting a symmetrized version of the experiment, still violating the CHSH inequality but now also obeying the marginal law, for which we provide a full quantum modeling in Hilbert space, using a singlet state and suitably chosen product measurements. We also address some of the criticisms that have been recently directed at experiments of this kind, according to which they would not highlight the presence of genuine forms of entanglement. We explain that these criticisms are based on a view of entanglement that is too restrictive, thus unable to capture all possible ways physical and conceptual entities can connect and form systems behaving as a whole. We also provide an example of a mechanical model showing that the violations of the marginal law and Bell inequalities are generally to be associated with different mechanisms.
\end{abstract}

\section{Introduction\label{intro}}

Following its theoretical discovery, by Einstein, Podolsky, Rosen \cite{epr1935} and Schr{\" o}dinger \cite{s1935a, s1935b}, entanglement has been extensively studied in physics, both theoretically and experimentally, and is today a well-established phenomenon, thanks also to the work of Bell \cite{bell1964,bell1966}, who showed that its presence in a composite system has experimental consequences: the violation of specific inequalities, involving observable quantities, that today bear his name. More precisely, the latter express constraints on the correlations between pairs of outcomes of jointly measurable observables, that must be satisfied by isolated (experimentally separated) systems. 

In recent times, the quantum formalism has also been successfully applied to model human cognition (see, e.g., 
\cite{a2009a,pb2009,k2010,bpft2011,bb2012,hk2013,pb2013,wbap2013,abgs2013,ags2013,ast2014,asdb2015b} and references therein), so that the question of the identification of entanglement in semantic spaces, in addition to physical spaces, also arose in a natural way. This especially if one considers an operational-realistic approach to cognition, like the one our group developed in Brussels \cite{ass2016b}, where conceptual entities, like the physical ones, are described in terms of basic notions like `state', `context' and `property'. This means that a meaningful parallel can be traced between `physics laboratories' and `psychological laboratories', as also in the latter one can define measurements to be performed on situations that are prepared in specific states, collect empirical data from the responses of the participants and deduce the probabilities associated to their responses, interpreted as measurement outcomes. The analysis of the latter by means of the formalism of quantum theory (or of more general quantum-like formalisms) is in fact the very object of study of the emerging field of `quantum cognition'.

In a nutshell, entanglement in physics is the experimental evidence that a composite system can be put in states (called entangled states) such that although its components are possibly separated by arbitrary spatial distances, they can nevertheless remain connected in ways that when measurements are performed on the different components (in sequence or in a coincident way), they will necessarily influence each other. In other words, entanglement in physics is about the fact that spatial separation is not sufficient to achieve experimental separation. This was the ``mistake'' committed by Einstein, Podolsky and Rosen, when they stated their famous paradox \cite{epr1935}, as they precisely assumed that because of their spatial distance the components of a bipartite entity could be considered as not disturbing each other any more. But experiments indicated a very different state of affairs \cite{aspect1982a, aspect1983,aspect1982b,tittel1998,weihs1998,genovese2005,vienna2013,urbana2013,hensen-etal2015}: 
that entanglement is the expression of a mutual influence between the sub-systems of a composite system which is \emph{non-spatial} in nature, so that it cannot be eliminated by simply putting some spatial distance between the locations of the joint measurements, even if sufficient to avoid signaling. 

The role of entanglement in human cognition was firstly emphasized in \cite{nm2007, bkmm2008, bknm2009}, in the ambit of word associations. It was subsequently studied in \cite{as2011, as2014}, in relation to concept combinations, showing that the CHSH version of Bell's inequality can be significantly violated by the experimental data. In \cite{as2011}, the presence of entanglement was revealed by considering two concepts, \emph{Animal} and \emph{Acts}, in the context of their combination \emph{The Animal Acts}. Participants in the experiment were asked to select pairs of exemplars for these two concepts, as good examples of their combination. Considering two couples of exemplars for each concept (\emph{Horse} and \emph{Bear}, then \emph{Tiger} and \emph{Cat}, for \emph{Animal}, defining measurements $A$ and $A'$, respectively; \emph{Growls} and \emph{Whinnies}, then \emph{Snorts} and \emph{Meows}, for \emph{Acts}, defining measurements $B$ and $B'$, respectively), it was then possible to define four joint measurements $AB$, $A'B$, $AB'$ and $A'B'$, and use the statistics of their outcomes to test the CHSH inequality: 
\begin{equation}\label{chsh}
|S|\le 2, \quad S\equiv E(A,B)-E(A,B')+E(A',B)+E(A',B'),
\end{equation}
which was shown to be violated with value $|S|=2.4197$. In (\ref{chsh}), $E(A,B)$ is the expectation value for the joint measurement $AB$, given by:
$E(A,B)=p(A_1,B_1)-p(A_1,B_2)-p(A_2,B_1)+p(A_2,B_2)$, where $p(A_1,B_1)$ is the probability for obtaining the pair of outcomes $A_1=$ \emph{Horse} and $B_1 =$ \emph{Growls}, corresponding to the combination \emph{The Horse Growls}, and similarly for the other probabilities and joint measurements. 

\begin{figure}[htbp]
\begin{center}
\includegraphics[width=17cm]{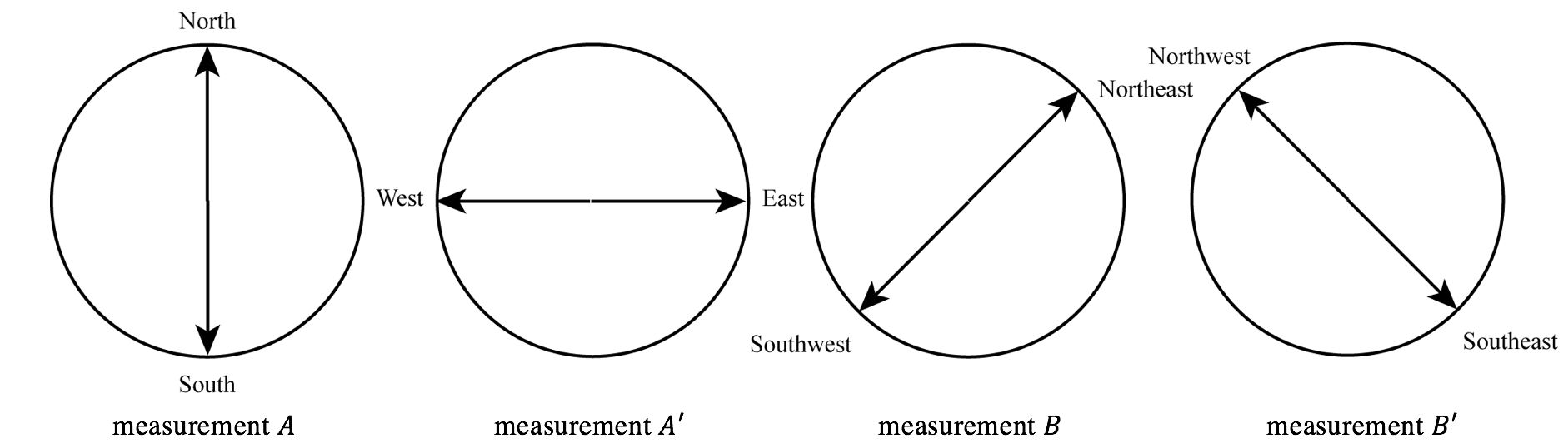}
\caption{A graphical representation of the two outcomes for the four measurements $A$, $A'$, $B$ and $B'$, performed on the two elements of the\emph{Two Different Wind Directions} composite conceptual entity.}
\label{Figure1}
\end{center}
\end{figure}

\begin{figure}[htbp]
\begin{center}
\includegraphics[width=16cm]{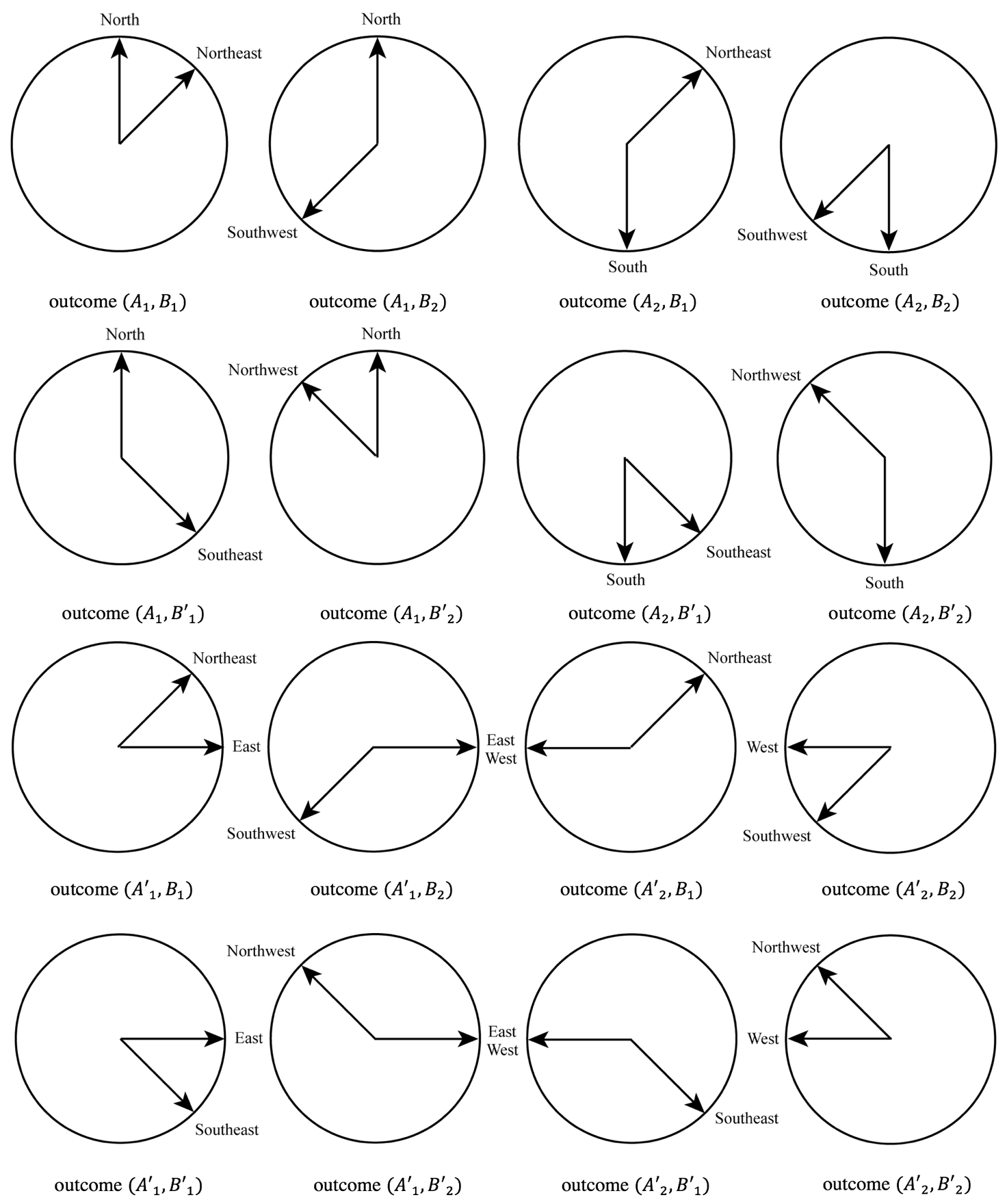}
\caption{A graphical representation of the four possible outcomes of the joint measurements $AB$, $AB'$, $A'B$ and $A'B'$, performed on the \emph{Two Different Wind Directions} conceptual entity.}
\label{Figure2}
\end{center}
\end{figure}

More recently, the conceptual combination \emph{Two Different Wind Directions} was also studied, asking again participants to provide good examples of it, choosing among pairs of different directions that were proposed to them. More specifically (see \cite{ass2017} for more details about the experiment, which involved 85 individuals), measurements $A$ and $A'$, on the first {\it One Wind Direction} conceptual element, were taken to have the outcomes $A_1$ = {\it North}, $A_2$ = {\it South} and $A'_1$ = {\it East} and $A'_2$ = {\it West}, respectively. On the other hand, measurements $B$ and $B'$ on the {\it Other Wind Direction} conceptual element, were taken to have the outcomes $B_1$ = {\it Northeast}, $B_2$ = {\it Southwest} and $B'_1$ = {\it Southeast}, $B'_2$ = {\it Northwest}, respectively (see Fig.~\ref{Figure1}).
For the four joint measurements $AB$, $A'B$, $AB'$ and $A'B'$, we thus obtain the following combined outcomes (see Fig.~\ref{Figure2}). 
For $AB$: $(A_1,B_1)$ = ({\it North}, {\it Northeast}), $(A_1,B_2)$ = ({\it North}, {\it Southwest}), $(A_2,B_1)$ = ({\it South}, {\it Northeast}) and $(A_2,B_2)$ = ({\it South}, {\it Southwest}). 
For $AB'$: $(A_1,B'_1)$ = ({\it North}, {\it Southeast}), $(A_1,B'_2)$ = ({\it North}, {\it Northwest}), $(A_2,B'_1)$ = ({\it South}, {\it Southeast}) and $(A_2,B'_2)$ = ({\it South}, {\it Northwest}). 
For $A'B$: $(A'_1,B_1)$ = ({\it East}, {\it Northeast}), $(A'_1,B_2)$ = ({\it East}, {\it Southwest}), $(A'_2,B_1)$ = ({\it West}, {\it Northeast}) and $(A'_2,B_2)$ = ({\it West}, {\it Southwest}). 
For $A'B'$: $(A'_1,B'_1)$ = ({\it East}, {\it Southeast}), $(A'_1,B'_2)$ = ({\it East}, {\it Northwest}), $(A'_2,B'_1)$ = ({\it West}, {\it Southeast}) and $(A'_2,B'_2)$ = ({\it West}, {\it Northwest}).
Considering the obtained statistics of outcomes \cite{ass2017}: 
\begin{eqnarray}
&&p(A_1,B_1)=0.13,\quad p(A_1,B_2)=0.55,\quad p(A_2,B_1)= 0.25,\quad p(A_2,B_2)=0.07 \nonumber\\
&&p(A_1,B'_1)=0.47,\quad p(A_1,B'_2)= 0.12,\quad p(A_2,B'_1)= 0.06,\quad p(A_2,B'_2)=0.35 \nonumber\\
&&p(A'_1,B_1)=0.13,\quad p(A'_1,B_2)=0.38,\quad p(A'_2,B_1)= 0.42,\quad p(A'_2,B_2)=0.07 \nonumber\\
&&p(A'_1,B'_1)= 0.09,\quad p(A'_1,B'_2)= 0.44,\quad p(A'_2,B'_1)=0.38,\quad p(A'_2,B'_2)=0.09,
\label{data}
\end{eqnarray}
the CHSH inequality was shown to be violated with value $|S|=2.47$.

It is important at this point to emphasize a difference between the violation of the CHSH inequality in the above two psychological experiments and in typical Bell-test experiments in physics. To this end, let us recall that the CHSH inequality is typically derived by assuming the existence of a general local hidden-variable theory, allowing for a classical description of the physical reality resulting from the specification of the unknown (hidden) variables that are to be added to the quantum state vectors. From an operational point of view, this is about assuming the experimental separation of the two sub-systems, which from a statistical point of view means that the outcomes of the measurements performed on the sub-systems are statistically independent from one another. 

An immediate consequence of this is that experimentally separated sub-systems will not only obey the CHSH inequality, but also the so-called `marginal law' (also called `marginal selectivity', `no-signaling condition', or `no-disturbance principle'). The latter states that, when we sum over the outcomes of one measurement, the obtained probabilities are independent of it. More specifically, the marginal law is the requirement that: 
\begin{eqnarray}\label{marginal}
&&\sum_{j=1,2}p(A_i,B_j) = \sum_{j=1,2}p(A_i,B'_j),\quad \sum_{j=1,2}p(A_j,B_i) = \sum_{j=1,2}p(A'_j,B_i),\quad i=1,2.\nonumber\\
&&\sum_{j=1,2}p(A'_i,B_j) = \sum_{j=1,2}p(A'_i,B'_j),\quad \sum_{j=1,2}p(A_j,B'_i) = \sum_{j=1,2}p(A'_j,B'_i),\quad i=1,2.
\end{eqnarray}

The above eight equalities are automatically satisfied in quantum theory, if the four joint measurements are described as product measurements, i.e., are associated with the tensor product observables: $A\otimes B$, $A'\otimes B$, $A\otimes B'$ and $A'\otimes B'$ (using here the same symbols to denote measurements and observables). Note that for measurements of this kind, the obedience to the marginal law is necessary to exclude mechanisms by which information could be transmitted backward in time \cite{e1978}. In that respect, it is interesting to observe that Aspect noticed in his PhD thesis \cite{aspect1983} that the marginal law is not always conserved in Bell-test experiments, although this was typically attributed to defects of the experimental setting in the preparation and/or registration (see also \cite{ak2007}). A different possible explanation, on which we will say something more in Sec.~\ref{conclusions}, is that measurements in Bell-type experiments may not always be characterized as product measurements.

Now, in experiments with the two composite conceptual entities \emph{The Animal Acts} and \emph{Two Different Wind Directions}, the marginal law is typically violated.\footnote{This of course is not in conflict with special relativity, as the notion of `space-like interval' does not apply to concepts.} Indeed, in view of (\ref{data}), we have:\footnote{Note that the violation is significant only in a few cases, as it may be evidenced by performing a Student t-test. More precisely, by computing a two tails two paired samples t-test for means of $p(A_1,B_1)+p(A_1,B_2)$ and $p(A_1,B'_1)+p(A_1,B'_2)$, one finds a p-value $p(df=84)=0.17$. Analogously, comparison between $p(A_2,B_1)+p(A_2,B_2)$ and $p(A_2,B'_1)+p(A_2,B'_2)$ gives $p(df=84)=0.17$; comparison between $p(A'_1,B_1)+p(A'_1,B_2)$ and $p(A'_1,B'_1)+p(A'_1,B'_2)$ gives $p(df=84)=0.76$; comparison between $p(A'_2,B_1)+p(A'_2,B_2)$ and $p(A'_2,B'_1)+p(A'_2,B'_2)$ gives $p(df=84)=0.76$; comparison between $p(A_1,B'_1)+p(A_2,B'_1)$ and $p(A'_1,B'_1)+p(A'_2,B'_1)$ gives $p(df=84)=0.47$; comparison between $p(A_1,B'_2)+p(A_2,B'_2)$ and $p(A'_1,B'_2)+p(A'_2,B'_2)$ gives $p(df=84)=0.47$. Finally, comparisons between $p(A_1,B_1)+p(A_2,B_1)$ and $p(A'_1,B_1)+p(A'_2,B_1)$, and between $p(A_1,B_2)+p(A_2,B_2)$ and $p(A'_1,B_2)+p(A'_2,B_2)$, gives a relatively small p-value of $p(df=84)=0.02$.}
\begin{eqnarray}
&&p(A_1,B_1)+p(A_1,B_2)=0.68 \ne 0.59=p(A_1,B'_1)+p(A_1,B'_2) \nonumber\\
&&p(A_2,B_1)+p(A_2,B_2)=0.32 \ne 0.41=p(A_2,B'_1)+p(A_2,B'_2) \nonumber\\
&&p(A_1,B_1)+p(A_2,B_1)=0.38 \ne 0.55=p(A'_1,B_1)+p(A'_2,B_1) \nonumber\\
&&p(A_1,B_2)+p(A_2,B_2)=0.62 \ne 0.45=p(A'_1,B_2)+p(A'_2,B_2), \nonumber\\
&&p(A'_1,B_1)+p(A'_1,B_2)=0.51 \ne 0.53=p(A'_1,B'_1)+p(A'_1,B'_2) \nonumber\\
&&p(A'_2,B_1)+p(A'_2,B_2)=0.49 \ne 0.47=p(A'_2,B'_1)+p(A'_2,B'_2) \nonumber\\
&&p(A_1,B'_1)+p(A_2,B'_1)=0.53 \ne 0.47=p(A'_1,B'_1)+p(A'_2,B'_1) \nonumber\\
&&p(A_1,B'_2)+p(A_2,B'_2)=0.47 \ne 0.53=p(A'_1,B'_2)+p(A'_2,B'_2).
\end{eqnarray}
A violation of the marginal law in situations where the CHSH inequality is violated is however to be expected. Indeed, the CHSH inequality is derived under the assumption that the sub-systems are experimentally separated. This automatically guarantees the preservation of the marginal law. On the other hand, if the sub-systems cease to be experimentally separated, this means that some kind of connection exists between them, which explains why correlations can be actualized in joint measurements that are able to violate the CHSH and other Bell-like inequalities. But the very existence of a connection of some kind implies that the marginal law needs not to be satisfied, as the condition for its derivation is not anymore fulfilled. Of course, this does not mean that the marginal law needs always  to be violated. In that respect, the situation of the standard quantum formalism appears to particular, i.e., the expression of a remarkable underlying symmetry, which however needs not to be present in all entangled systems. 

When the marginal law is violated, one cannot model the experimental probabilities using product measurements. However, this does not mean that the standard quantum formalism would be unable to model the data. To this aim, it is sufficient to introduce a refined scheme where not only entangled states, but also entangled (non-product) measurements can occur. A full quantum modeling can then be carried out, as demonstrated in \cite{as2014} for the \emph{The Animal Acts} situation. So, cognitive psychology experiments are in a sense able to exhibit a stronger form of entanglement in comparison to experiments with micro-physical entities. However, the question arises if psychological experiments disobeying the CHSH inequality, but obeying the marginal law, can be identified. This seems to be an important question in view of a generalized definition of contextuality proposed by some authors \cite{dk2014,dkczj2016} (we will come back to this in Sec.~\ref{discussion}). 

It is one of the purposes of the present article to show that such experiment exists and is easy to define, which we will do in Sec.~\ref{version}. Then, in Sec.~\ref{quantummodel}, we provide an explicit and faithful quantum theoretical modeling of the experimental data. Since the latter obey the marginal law, the representation only uses product measurements and an entangled (singlet) state, hence showing a remarkable similarity with the typical description of Bell-test experiments in physics. In Sec.~\ref{discussion}, we provide answers to criticisms that have been made about the pertinence of our approach \cite{dk2014,dkczj2016}, also using the example of the non-uniform rigid rod mechanical model, analyzed in Sec.~\ref{appendix}, as a way to make more evident our conclusions. Finally, in Sec.~\ref{conclusions}, we offer some concluding remarks.

\section{An experiment that preserves the marginal law \label{version}}

In the experiment described in \cite{ass2017}, the four coincidence measurements $AB$, $AB'$, $A'B$ and $A'B'$, performed one the conceptual entity \emph{Two Different Wind Directions} (see Fig.~\ref{Figure2}), were defined starting from the four one-entity measurements indicated in Fig.~\ref{Figure1}, each one corresponding to a different spatial axis. More precisely, $A$ was chosen to correspond to the {\it South}-{\it North} spatial axis, then $B$, $A'$ and $B'$ were clockwise rotated $45^\circ$, $90^\circ$ and $135^\circ$ with respect to $A$, respectively. The reason for this specific choice is that it corresponds to the four main axes associated with the so-called windrose (or mariner's compass rose), associated with the eight traditional wind directions.\footnote{The name of the eight traditional wind directions are: Tramontane ({\it North}), Greco ({\it Northeast}), Levante ({\it East}), Sirocco ({\it Southeast}), Ostro ({\it South}), Libeccio ({\it Southwest}), Ponente ({\it West}) and Maestro ({\it Northwest}).} 

A `wind direction' is a different notion than a `space direction'. All space directions are in fact assumed to be equivalent for physical entities (isotropy of space), whereas wind directions are usually associated with very specific space directions, typically the cardinal directions, which resulted from our human experience on the surface of our planet. This means that an experiment about wind directions cannot exhibit the same level of symmetry of an experiment about space directions, as is clear that \emph{Wind}, as a concept, is perceived quite differently from \emph{Space}. However, Bell-test experiments in physics only deal with spatial orientations of the measuring apparatuses (like the Stern-Gerlach ones, in spin measurements), so one may wonder if the violation of the marginal law in our cognitive experiment with the \emph{Two Different Wind Directions} entity would not be the consequence of this inevitable symmetry breaking introduced by the biases associated with the {\it Wind} concept. 

Therefore, one may also wonder whether it is possible to design an experiment where the symmetry would be restored, i.e., such that the marginal law would be recovered, with the CHSH inequality being however still violated. The answer is affirmative, as we are now going to show. For this, we observe that we humans associate different meanings to \emph{South} and \emph{North}, as well as to \emph{East} and \emph{West}. Hence, we can expect that an experiment uniformly mixing \emph{South} with \emph{North} and \emph{East} with \emph{West} (and consequently \emph{Southwest} with \emph{Northeast} and \emph{Southeast} with \emph{Northwest}) should be able to bring the notion of `wind directions' closer to that of `space directions'. So, we consider an experiment with the following protocol. 

We start by flipping a coin. If heads is obtained, we perform the four measurements $AB$, $AB'$, $A'B$ and $A'B'$. If tails is obtained, we perform these same measurements but with the axes of measurements $A$, $A'$, $B$ and $B'$ all rotated $180^\circ$. This corresponds to the situation described in Fig.~\ref{Figure1}, with the labels of the outcomes interchanged (note that the relative orientation of the different measurements remain unaffected, as they are all rotated of the same angle). The joint measurements associated with this $180^\circ$-rotated version of the experiment are thus the following. For $AB$: $(A_1,B_1)$ = ({\it South}, {\it Southwest}), $(A_1,B_2)$ = ({\it South}, {\it Northeast}), $(A_2,B_1)$ = ({\it North}, {\it Southwest}) and $(A_2,B_2)$ = ({\it North}, {\it Northeast}). For $AB'$: $(A_1,B'_1)$ = ({\it South}, {\it Northwest}), $(A_1,B'_2)$ = ({\it South}, {\it Southeast}), $(A_2,B'_1)$ = ({\it North}, {\it Northwest}) and $(A_2,B'_2)$ = ({\it North}, {\it Southeast}). For $A'B$: $(A'_1,B_1)$ = ({\it West}, {\it Southwest}), $(A'_1,B_2)$ = ({\it West}, {\it Northeast}), $(A'_2,B_1)$ = ({\it East}, {\it Southwest}) and $(A'_2,B_2)$ = ({\it East}, {\it Northeast}). For $A'B'$: $(A'_1,B'_1)$ = ({\it West}, {\it Northwest}), $(A'_1,B'_2)$ = ({\it West}, {\it Southeast}), $(A'_2,B'_1)$ = ({\it East}, {\it Northwest}) and $(A'_2,B'_2)$ = ({\it East}, {\it Southeast}).

Let us denote $p_{180}$ the probabilities associated with the $180^\circ$-rotated measurements. Interestingly, we don't really need to perform these rotated measurements to obtain their joint probabilities. Indeed, we only have for this to reallocate the data (\ref{data}) collected for the unrotated measurements, under the assumption that participants, when performing the rotated measurements, will remain consistent in their choices with respect to how they have selected outcomes in the unrotated ones. Then, we have: $p_{\rm 180^\circ}(A_1,B_1) = p(A_2,B_2)$, $p_{180}(A_1,B_2) = p(A_2,B_1)$, $p_{180}(A_1,B'_1) = p(A_2,B'_2)$, $p_{180}(A_1,B'_2) = p(A_2,B'_1)$, $p_{180}(A'_1,B_1) = p(A'_2,B_2)$, $p_{180}(A'_1,B_2) = p(A'_2,B_1)$, $p_{180}(A'_1,B'_1) = p(A'_2,B'_2)$, $p_{180}(A'_1,B'_2) = p(A'_2,B'_1)$. Similarly: $p_{180}(A_2,B_2) = p(A_1,B_1)$, $p_{180}(A_2,B_1) = p(A_1,B_2)$, $p_{180}(A_2,B'_1) = p(A_1,B'_2)$, $p_{180}(A_2,B'_2) = p(A_1,B'_1)$, $p_{180}(A'_2,B_1) = p(A'_1,B_2)$, $p_{180}(A'_2,B_2) = p(A'_1,B_1)$, $p_{180}(A'_2,B'_1) = p(A'_1,B'_2)$, $p_{180}(A'_2,B'_2) = p(A'_1,B'_1)$. So, the probability for, say, outcome $(A_1,B_1)$, in the mixed measurement where the rotated and unrotated situations are chosen randomly, is given by the uniform average $\bar{p}(A_1,B_1)=\frac{1}{2}[p(A_1,B_1)+p_{180}(A_1,B_1)]$, and similarly for the other outcomes. We thus obtain: 
\begin{eqnarray}
&\bar{p}(A_1,B_1)=\bar{p}(A_2,B_2) =\frac{1}{2}[p(A_1,B_1)+p(A_2,B_2)]=0.10 \nonumber\\
&\bar{p}(A_1,B_2)=\bar{p}(A_2,B_1)=\frac{1}{2}[p(A_1,B_2)+p(A_2,B_1)]=0.40 \nonumber\\
&\bar{p}(A_1,B'_1)=\bar{p}(A_2,B'_2)=\frac{1}{2}[p(A_1,B'_1)+p(A_2,B'_2)]=0.41 \nonumber\\
&\bar{p}(A_1,B'_2)=\bar{p}(A_2,B'_1)=\frac{1}{2}[p(A_1,B'_2)+p(A_2,B'_1)]=0.09 \nonumber\\
&\bar{p}(A'_1,B_1)=\bar{p}(A'_2,B_2)=\frac{1}{2}[p(A'_1,B_1)+p(A'_2,B_2)]=0.10 \nonumber\\
&\bar{p}(A'_1,B_2)=\bar{p}(A'_2,B_1)=\frac{1}{2}[p(A'_1,B_2)+p(A'_2,B_1)]=0.40 \nonumber\\
&\bar{p}(A'_1,B'_1)=\bar{p}(A'_2,B'_2)=\frac{1}{2}[p(A'_1,B'_1)+p(A'_2,B'_2)]=0.09 \nonumber\\
&\bar{p}(A'_1,B'_2)=\bar{p}(A'_2,B'_1)=\frac{1}{2}[p(A'_1,B'_2)+p(A'_2,B'_1)]=0.41.
\end{eqnarray}
The marginal law is now satisfied, as we have:
\begin{eqnarray}
&&\bar{p}(A_1,B_1)+\bar{p}(A_1,B_2)=0.5=\bar{p}(A_1,B'_1)+\bar{p}(A_1,B'_2) \nonumber\\
&&\bar{p}(A_2,B_1)+\bar{p}(A_2,B_2)=0.5=\bar{p}(A_2,B'_1)+\bar{p}(A_2,B'_2) \nonumber\\
&&\bar{p}(A_1,B_1)+\bar{p}(A_2,B_1)=0.5=\bar{p}(A'_1,B_1)+\bar{p}(A'_2,B_1) \nonumber\\
&&\bar{p}(A_1,B_2)+\bar{p}(A_2,B_2)=0.5=\bar{p}(A'_1,B_2)+\bar{p}(A'_2,B_2), \nonumber\\
&&\bar{p}(A'_1,B_1)+\bar{p}(A'_1,B_2)=0.5=\bar{p}(A'_1,B'_1)+\bar{p}(A'_1,B'_2) \nonumber\\
&&\bar{p}(A'_2,B_1)+\bar{p}(A'_2,B_2)=0.5=\bar{p}(A'_2,B'_1)+\bar{p}(A'_2,B'_2) \nonumber\\
&&\bar{p}(A_1,B'_1)+\bar{p}(A_2,B'_1)=0.5=\bar{p}(A'_1,B'_1)+\bar{p}(A'_2,B'_1) \nonumber\\
&&\bar{p}(A_1,B'_2)+\bar{p}(A_2,B'_2)=0.5=\bar{p}(A'_1,B'_2)+\bar{p}(A'_2,B'_2).
\end{eqnarray}
Also, for the expectation values, we obtain:
\begin{eqnarray}
&&\bar{E}(A,B)=\bar{p}(A_1,B_1)-\bar{p}(A_1,B_2)-\bar{p}(A_2,B_1)+\bar{p}(A_2,B_2)=-0.60 \nonumber \\
&&\bar{E}(A,B')=\bar{p}(A_1,B'_1)-\bar{p}(A_1,B'_2)-\bar{p}(A_2,B'_1)+\bar{p}(A_2,B'_2)=+0.65 \nonumber \\
&&\bar{E}(A',B)=\bar{p}(A'_1,B_1)-\bar{p}(A'_1,B_2)-\bar{p}(A'_2,B_1)+\bar{p}(A'_2,B_2)=-0.60 \nonumber \\
&&\bar{E}(A',B')=\bar{p}(A'_1,B'_1)-\bar{p}(A'_1,B'_2)-\bar{p}(A'_2,B'_1)+\bar{p}(A'_2,B'_2)=-0.62,
\end{eqnarray}
so that the CHSH inequality (\ref{chsh}) is still violated, as we have: 
\begin{equation}\label{chsh2}
|\bar{E}(A,B)-\bar{E}(A,B')+\bar{E}(A',B)+\bar{E}(A',B')|= 2.47.
\end{equation}

What we have described is an experimental situation where there is preservation of the marginal law and, at the same time, violation of the CHSH inequality, with same magnitude of violation of typical Bell-test experiments in physics (see also the discussion in \cite{ass2017}).\footnote{The violation is statistically significant, as in a one sample t-test against the value 2 we obtain: $p(df=169)=0.01 < 0.05$. Note that since the unrotated experiment involved 85 participants \cite{ass2017}, this mixed version of the measurement provides a total of 170 estimations. } However, we can even go one step further, and define mixed joint measurements whose probabilities and expectation values present the same level of symmetry of those theoretically predicted by quantum theory for coincidence spin measurements on pairs of spin-${1\over 2}$ entities in singlet states. For this, instead of flipping a coin, we now roll a die with eight faces (octahedron), reporting on them the following angles: $0^\circ$, $45^\circ$, $90^\circ$, $135^\circ$, $180^\circ$, $225^\circ$, $270^\circ$ and $315^\circ$. 

If the $0^\circ$-face is obtained, we simply perform the four measurements $AB$, $AB'$, $A'B$ and $A'B'$. If the $45^\circ$-face is obtained, we perform these same four joint measurements, but with the axes of the individual measurements $A$, $A'$, $B$ and $B'$ all rotated $45^\circ$ clockwise. This means that the outcomes of the $45^\circ$-rotated joint measurement $AB$ are: $(A_1,B_1)$ = ({\it Northeast}, {\it East}), $(A_1,B_2)$ = ({\it Northeast}, {\it West}), $(A_2,B_1)$ = ({\it Southwest}, {\it East}) and $(A_2,B_2)$ = ({\it Southwest}, {\it West}). Note that these outcomes correspond to the outcomes of the joint measurement $A'B$ of the unrotated experiment, with the order of the outcomes having different indexes interchanged. This means that, assuming as before that participants always choose in a consistent way, we can write, with obvious notation: $p_{45}(A_1,B_1)=p(A'_1,B_1)$, $p_{45}(A_2,B_2)=p(A'_2,B_2)$, $p_{45}(A_1,B_2)=p(A'_2,B_1)$ and $p_{45}(A_2,B_1)=p(A'_1,B_2)$. And of course we can proceed in a similar way for the measurements $AB'$, $A'B$ and $A'B'$ of the $45^\circ$-rotated situation, looking for the correspondences with the outcomes of the unrotated situation.

The reasoning for the measurements with the other rotation angles is analogous. For the $90^\circ$ rotation we have, for the joint measurement $AB$: 
$(A_1,B_1)$ = ({\it East}, {\it Southeast}), $(A_1,B_2)$ = ({\it East}, {\it Northwest}), $(A_2,B_1)$ = ({\it West}, {\it Southeast}) and $(A_2,B_2)$ = ({\it West}, {\it Northwest}). These outcomes correspond to those of the unrotated joint measurement $A'B'$, in exactly the same order, so that we have: $p_{90}(A_1,B_1)=p(A'_1,B'_1)$, $p_{90}(A_2,B_2)=p(A'_2,B'_2)$, $p_{90}(A_1,B_2)=p(A'_1,B'_2)$ and $p_{90}(A_2,B_1)=p(A'_2,B'_1)$.
Then we proceed similarly for the measurements $AB'$, $A'B$ and $A'B'$ of the $90^\circ$-rotated situation, looking again for the correspondences with the outcomes of the unrotated situation.

For the $135^\circ$-rotated situation, we have for measurement $AB$: $(A_1,B_1)$ = ({\it Southeast}, {\it South}), $(A_1,B_2)$ = ({\it Southeast}, {\it North}), $(A_2,B_1)$ = ({\it Northwest}, {\it South}) and $(A_2,B_2)$ = ({\it Northwest}, {\it North}). These outcomes correspond to those of the unrotated $AB'$ measurement, although in different order. More precisely: $p_{135}(A_1,B_1)=p(A_2,B'_1)$, $p_{135}(A_2,B_2)=p(A_1,B'_2)$, $p_{135}(A_1,B_2)=p(A_1,B'_1)$ and $p_{135}(A_2,B_1)=p(A_2,B'_2)$. And we proceed in a similar way for the measurements $AB'$, $A'B$ and $A'B'$ of the $135^\circ$-rotated situation, looking for the correspondences with the outcomes of the unrotated situation.

The recollection of the probabilistic data for the remaining rotated measurements proceeds analogously. We thus have a mixture of eight rotated sets of joint measurements $AB$, $AB'$, $A'B$, and $A'B'$. Since the unrotated one involved 85 participants \cite{ass2017}, we now have a total of 680 estimations. The probability for measurement $AB$, to yield outcome $(A_1,B_1)$, is then given by the uniform average: 
\begin{eqnarray}
\lefteqn{\bar{p}(A_1,B_1)=\frac{1}{8}[ p(A_1,B_1)+p(A'_1,B_1)+p(A'_1,B'_1)+p(A_2,B'_1)} \nonumber \\ 
&\quad +\, p(A_2,B_2)+p(A'_2,B_2)+p(A'_2,B'_2)+p(A_1,B'_2) ]=0.095.
\end{eqnarray}
Proceeding in the same way with the other outcome probabilities and joint measurements, we obtain: 
\begin{eqnarray}
&&\bar{p}(A_1,B_1)=0.095,\quad \bar{p}(A_1,B_2)=0.405,\quad \bar{p}(A_2,B_1)=0.405,\quad \bar{p}(A_2,B_2)= 0.095\nonumber\\
&&\bar{p}(A_1,B'_1)=0.405,\quad \bar{p}(A_1,B'_2)= 0.095,\quad \bar{p}(A_2,B'_1)=0.095,\quad \bar{p}(A_2,B'_2)=0.405 \nonumber\\
&&\bar{p}(A'_1,B_1)=0.095,\quad \bar{p}(A'_1,B_2)=0.405,\quad \bar{p}(A'_2,B_1)=0.405 ,\quad \bar{p}(A'_2,B_2)=0.095 \nonumber\\
&&\bar{p}(A'_1,B'_1)=0.095,\quad \bar{p}(A'_1,B'_2)= 0.405,\quad \bar{p}(A'_2,B'_1)=0.405,\quad \bar{p}(A'_2,B'_2)=0.095.
\label{data2}
\end{eqnarray}
The corresponding expectation values are:
\begin{eqnarray}
&&\bar{E}(A,B)=\bar{p}(A_1,B_1)-\bar{p}(A_1,B_2)-\bar{p}(A_2,B_1)+\bar{p}(A_2,B_2)=-0.62 \nonumber \\
&&\bar{E}(A,B')=\bar{p}(A_1,B'_1)-\bar{p}(A_1,B'_2)-\bar{p}(A_2,B'_1)+\bar{p}(A_2,B'_2)=0.62 \nonumber \\
&&\bar{E}(A',B)=\bar{p}(A'_1,B_1)-\bar{p}(A'_1,B_2)-\bar{p}(A'_2,B_1)+\bar{p}(A'_2,B_2)=-0.62 \nonumber \\
&&\bar{E}(A',B')=\bar{p}(A'_1,B'_1)-\bar{p}(A'_1,B'_2)-\bar{p}(A'_2,B'_1)+\bar{p}(A'_2,B'_2)=-0.62,
\end{eqnarray}
hence, the CHSH inequality (\ref{chsh}) is violated, exactly as in (\ref{chsh2}).\footnote{The violation is statistically significant: $p(df=679) \ll 0.05$, in a one sample t-test against the constant value 2.} Also, as it was the case in the previous mixed experiment, one can easily check that the marginal law is obeyed. Therefore, we have now described an experimental situation where the symmetry breaking between `space directions' and `wind directions' has been eliminated.

\section{A Hilbert space quantum model\label{quantummodel}}

In this section, we provide an explicit quantum theoretical representation, in complex Hilbert space, for the statistical data (\ref{data2}). Since the latter come from a fully symmetrized version of the measurements presented in \cite{ass2017}, we will now interpret them as data modeling the conceptual combination {\it Two Different Space Directions}, instead of {\it Two Different Wind Directions}. To construct the quantum model, we observe that since joint measurements have four outcomes, the Hilbert space to consider is ${\mathbb C}^{4}$. Here we take advantage of the canonical isomorphism between ${\mathbb C}^{4}$ and ${\mathbb C}^{2} \otimes {\mathbb C}^{2}$, so that we can consider as a basis of the latter the four tensor product unit vectors: $|f_1\rangle \otimes |f_1\rangle$, $|f_1\rangle\otimes |f_2\rangle$, $|f_2\rangle\otimes |f_1\rangle$ and $|f_2\rangle\otimes |f_2\rangle$, where the two orthonormal vectors $|f_1\rangle$ and $|f_2\rangle$ form a basis of ${\mathbb C}^{2}$. If we take them to be canonical:
\begin{equation}
|f_1\rangle=\left[ \begin{array}{c}
1\\ 0
\end{array} \right],\quad 
|f_2\rangle=\left[ \begin{array}{c}
0\\ 1
\end{array} \right],
\end{equation}
then we also have: 
\begin{equation}
|f_1\rangle \otimes |f_1\rangle=\left[ \begin{array}{c}
1\\ 0\\ 0\\ 0
\end{array} \right],\quad 
|f_1\rangle\otimes |f_2\rangle=\left[ \begin{array}{c}
0\\ 1\\ 0\\ 0
\end{array} \right],\quad
|f_2\rangle\otimes |f_1\rangle=\left[ \begin{array}{c}
0\\ 0\\ 1\\ 0
\end{array} \right],\quad
|f_2\rangle\otimes |f_2\rangle=\left[ \begin{array}{c}
0\\ 0\\ 0\\ 1
\end{array} \right].
\end{equation}

For the initial state of the {\it Two Different Space Directions} conceptual entity, we choose the rotational invariant singlet state:
\begin{equation}
|\Psi\rangle = \frac{1}{\sqrt{2}}(|f_1\rangle \otimes |f_2\rangle - |f_2\rangle \otimes |f_1\rangle)=\frac{1}{\sqrt{2}}\left[\!\!\! \begin{array}{r}
0\\ 1\\ -1\\ 0
\end{array} \right],
\label{singlet}
\end{equation}
and we represent the four joint measurements $AB$, $AB'$, $A'B$ and $A'B'$ by means of the product spin-observables $(\boldsymbol{\sigma}\cdot {\bf a}) \otimes (\boldsymbol{\sigma}\cdot {\bf b})$, $(\boldsymbol{\sigma}\cdot {\bf a}) \otimes (\boldsymbol{\sigma}\cdot {\bf b'})$, $(\boldsymbol{\sigma}\cdot {\bf a'}) \otimes (\boldsymbol{\sigma}\cdot {\bf b})$ and $(\boldsymbol{\sigma}\cdot {\bf a'}) \otimes (\boldsymbol{\sigma}\cdot {\bf b'})$, respectively, where ${\bf a}$, ${\bf a'}$, ${\bf b}$ and ${\bf b'}$ are unit vectors describing space directions, and $\boldsymbol{\sigma}$ is a vector whose components are the three Pauli's matrices:
\begin{equation}
\sigma_x=
\left[ \begin{array}{cc}
0 & 1 \\
1 & 0
\end{array} \right],
\quad 
\sigma_y=
\left[ \begin{array}{cc}
0 & -i \\
i & 0
\end{array} \right],
\quad
\sigma_z=
\left[ \begin{array}{cc}
1 & 0 \\
0 & -1
\end{array} \right].
\end{equation}

Introducing the spherical coordinates: 
\begin{eqnarray}
{\bf a}= \left[\begin{array}{c}
\sin\theta_{a}\cos\phi_{a}\\ \sin\theta_{a}\sin\phi_{a}\\ \cos\theta_{a}
\end{array} \right],\ 
{\bf b}= \left[\begin{array}{c}
\sin\theta_{b}\cos\phi_{b}\\ \sin\theta_{b}\sin\phi_{b}\\ \cos\theta_{b}
\end{array} \right],\ 
{\bf a'}= \left[\begin{array}{c}
\sin\theta_{a'}\cos\phi_{a'}\\ \sin\theta_{a'}\sin\phi_{a'}\\ \cos\theta_{a'}
\end{array} \right],\
{\bf b'} =\left[\begin{array}{c}
\sin\theta_{b'}\cos\phi_{b'}\\ \sin\theta_{b'}\sin\phi_{b'}\\ \cos\theta_{b'}
\end{array} \right],
\label{unitvectors}
\end{eqnarray}
the eigenvectors of $(\boldsymbol{\sigma}\cdot {\bf a}) \otimes (\boldsymbol{\sigma}\cdot {\bf b})$ can be written as: 
\begin{eqnarray}
&&|A_{1}B_{1}\rangle=\left[ \begin{array}{l}
\cos\frac{\theta_a}{2} e^{-i {\phi_a \over 2}} \\
\sin\frac{\theta_a}{2} e^{i {\phi_a \over 2}}
\end{array} \right]
\otimes
\left[ \begin{array}{l}
\cos\frac{\theta_b}{2} e^{-i {\phi_b \over 2}} \\
\sin\frac{\theta_b}{2} e^{i {\phi_b \over 2}}
\end{array} \right]=
\left[ \begin{array}{l}
\cos\frac{\theta_a}{2} \cos\frac{\theta_b}{2}e^{-\frac{i}{2} (\phi_a+\phi_b)} \\
\cos\frac{\theta_a}{2} \sin\frac{\theta_b}{2}e^{\frac{i}{2} (\phi_b-\phi_a)} \\
\sin\frac{\theta_a}{2} \cos\frac{\theta_b}{2}e^{-\frac{i}{2} (\phi_b-\phi_a)} \\
\sin\frac{\theta_a}{2} \sin\frac{\theta_b}{2}e^{\frac{i}{2} (\phi_a+\phi_b)} 
\end{array} \right]
\nonumber\\
&&|A_{1}B_{2}\rangle = \left[ \begin{array}{l}
\cos\frac{\theta_a}{2} e^{-i {\phi_a \over 2}} \\
\sin\frac{\theta_a}{2} e^{i {\phi_a \over 2}}
\end{array} \right]
\otimes
\left[ \begin{array}{l}
-\sin\frac{\theta_b}{2} e^{-i {\phi_b \over 2}} \\
\cos\frac{\theta_b}{2} e^{i {\phi_b \over 2}}
\end{array} \right]=
\left[ \begin{array}{l}
-\cos\frac{\theta_a}{2} \sin\frac{\theta_b}{2}e^{-\frac{i}{2} (\phi_a+\phi_b)} \\
\cos\frac{\theta_a}{2} \cos\frac{\theta_b}{2}e^{\frac{i}{2} (\phi_b-\phi_a)} \\
-\sin\frac{\theta_a}{2} \sin\frac{\theta_b}{2}e^{-\frac{i}{2} (\phi_b-\phi_a)} \\
\sin\frac{\theta_a}{2} \cos\frac{\theta_b}{2}e^{\frac{i}{2} (\phi_a+\phi_b)} 
\end{array} \right]
\nonumber\\
&&|A_{2}B_{1}\rangle = \left[ \begin{array}{l}
-\sin\frac{\theta_a}{2} e^{-i {\phi_a \over 2}} \\
\cos\frac{\theta_a}{2} e^{i {\phi_a \over 2}}
\end{array} \right]
\otimes
\left[ \begin{array}{l}
\cos\frac{\theta_b}{2} e^{-i {\phi_b \over 2}} \\
\sin\frac{\theta_b}{2} e^{i {\phi_b \over 2}}
\end{array} \right]=
\left[ \begin{array}{l}
-\sin\frac{\theta_a}{2} \cos\frac{\theta_b}{2}e^{-\frac{i}{2} (\phi_a+\phi_b)} \\
-\sin\frac{\theta_a}{2} \sin\frac{\theta_b}{2}e^{\frac{i}{2} (\phi_b-\phi_a)} \\
\cos\frac{\theta_a}{2} \cos\frac{\theta_b}{2}e^{-\frac{i}{2} (\phi_b-\phi_a)} \\
\cos\frac{\theta_a}{2} \sin\frac{\theta_b}{2}e^{\frac{i}{2} (\phi_a+\phi_b)} 
\end{array} \right]
\nonumber\\
&&|A_{2}B_{2}\rangle = \left[ \begin{array}{l}
-\sin\frac{\theta_a}{2} e^{-i {\phi_a \over 2}} \\
\cos\frac{\theta_a}{2} e^{i {\phi_a \over 2}}
\end{array} \right]
\otimes
\left[ \begin{array}{l}
-\sin\frac{\theta_b}{2} e^{-i {\phi_b \over 2}} \\
\cos\frac{\theta_b}{2} e^{i {\phi_b \over 2}}
\end{array} \right]=
\left[ \begin{array}{l}
\sin\frac{\theta_a}{2} \sin\frac{\theta_b}{2}e^{-\frac{i}{2} (\phi_a+\phi_b)} \\
-\sin\frac{\theta_a}{2} \cos\frac{\theta_b}{2}e^{\frac{i}{2} (\phi_b-\phi_a)} \\
-\cos\frac{\theta_a}{2} \sin\frac{\theta_b}{2}e^{-\frac{i}{2} (\phi_b-\phi_a)} \\
\cos\frac{\theta_a}{2} \cos\frac{\theta_b}{2}e^{\frac{i}{2} (\phi_a+\phi_b)} 
\end{array} \right].\label{eigenAB}
\end{eqnarray}

We want the probabilities predicted by the Born rule to be equal to the experimental probabilities (\ref{data2}). We have: 
\begin{eqnarray} 
&&|\langle A_{1}B_{1}|\Psi\rangle|^{2}={1\over 2}| \cos\frac{\theta_a}{2} \sin\frac{\theta_b}{2}e^{\frac{i}{2} (\phi_b-\phi_a)} - \sin\frac{\theta_a}{2} \cos\frac{\theta_b}{2}e^{-\frac{i}{2} (\phi_b-\phi_a)}|^2\nonumber\\ 
&&={1\over 2}| \cos\frac{\theta_a}{2} \sin\frac{\theta_b}{2}(\cos {\phi_b-\phi_a\over 2}+i\sin {\phi_b-\phi_a\over 2}) - \sin\frac{\theta_a}{2} \cos\frac{\theta_b}{2}(\cos {\phi_b-\phi_a\over 2}-i\sin {\phi_b-\phi_a\over 2})|^2\nonumber\\ 
&&={1\over 2}| \cos {\phi_b-\phi_a\over 2}(\cos\frac{\theta_a}{2} \sin\frac{\theta_b}{2} - \sin\frac{\theta_a}{2} \cos\frac{\theta_b}{2}) +i\sin {\phi_b-\phi_a\over 2}(\cos\frac{\theta_a}{2} \sin\frac{\theta_b}{2} + \sin\frac{\theta_a}{2} \cos\frac{\theta_b}{2}) |^2\nonumber\\ 
&&= {1\over 2}| \cos {\phi_b-\phi_a\over 2}(-\sin {\theta_a-\theta_b\over 2}) +i\sin {\phi_b-\phi_a\over 2}(\sin {\theta_a+\theta_b\over 2}) |^2\nonumber\\ &&={1\over 2}\cos^2{\phi_b-\phi_a\over 2}\sin^2{\theta_a-\theta_b\over 2} + \sin^2 {\phi_b-\phi_a\over 2}\sin^2{\theta_a+\theta_b\over 2}\nonumber\\ &&={1\over 4}[1-\cos \theta_a\cos \theta_b -\sin \theta_a\sin \theta_b \cos(\phi_b-\phi_a) ],
\end{eqnarray}
\begin{eqnarray} 
&&|\langle A_{1}B_{2}|\Psi\rangle|^{2}={1\over 2}| \cos\frac{\theta_a}{2} \cos\frac{\theta_b}{2}e^{\frac{i}{2} (\phi_b-\phi_a)} + \sin\frac{\theta_a}{2} \sin\frac{\theta_b}{2}e^{-\frac{i}{2} (\phi_b-\phi_a)}|^2\nonumber\\ 
&&={1\over 2}| \cos\frac{\theta_a}{2} \cos\frac{\theta_b}{2}(\cos {\phi_b-\phi_a\over 2}+i\sin {\phi_b-\phi_a\over 2}) +\sin\frac{\theta_a}{2} \sin \frac{\theta_b}{2}(\cos {\phi_b-\phi_a\over 2}-i\sin {\phi_b-\phi_a\over 2})|^2\nonumber\\ 
&&={1\over 2}| \cos {\phi_b-\phi_a\over 2}(\cos\frac{\theta_a}{2} \cos\frac{\theta_b}{2} + \sin\frac{\theta_a}{2} \sin\frac{\theta_b}{2}) +i\sin {\phi_b-\phi_a\over 2}(\cos\frac{\theta_a}{2} \cos\frac{\theta_b}{2} - \sin\frac{\theta_a}{2} \sin\frac{\theta_b}{2}) |^2\nonumber\\ 
&&= {1\over 2}| \cos {\phi_b-\phi_a\over 2}\cos {\theta_a-\theta_b\over 2} +i\sin {\phi_b-\phi_a\over 2}\cos {\theta_a+\theta_b\over 2} |^2\nonumber\\ &&={1\over 2}\cos^2{\phi_b-\phi_a\over 2}\cos^2{\theta_a-\theta_b\over 2} + \sin^2 {\phi_b-\phi_a\over 2}\cos^2{\theta_a+\theta_b\over 2}\nonumber\\ 
&&={1\over 4}[1+\cos \theta_a\cos \theta_b +\sin \theta_a\sin \theta_b \cos(\phi_b-\phi_a) ]. \end{eqnarray}
We thus obtain the conditions: 
\begin{eqnarray}
&&|\langle A_{1}B_{1}|\Psi\rangle|^{2}= |\langle A_{2}B_{2}|\Psi\rangle|^{2}={1\over 4}[1-\cos \theta_a\cos \theta_b -\sin \theta_a\sin \theta_b \cos(\phi_b-\phi_a) ]=0.095\nonumber\\
&&|\langle A_{1}B_{2}|\Psi\rangle|^{2} = |\langle A_{2}B_{1}|\Psi\rangle|^{2}={1\over 4}[1+\cos \theta_a\cos \theta_b +\sin \theta_a\sin \theta_b \cos(\phi_b-\phi_a) ]=0.405,
\label{Born}
\end{eqnarray}
and similar expressions hold for measurements $AB'$, $A'B$ and $A'B'$. More precisely, for the eigenvectors of $(\boldsymbol{\sigma}\cdot {\bf a}) \otimes (\boldsymbol{\sigma}\cdot {\bf b'})$, we have the conditions: 
\begin{eqnarray}
&&|\langle A_{1}B'_{1}|\Psi\rangle|^{2}= |\langle A_{2}B'_{2}|\Psi\rangle|^{2}={1\over 4}[1-\cos \theta_a\cos \theta_{b'} -\sin \theta_a\sin \theta_{b'} \cos(\phi_{b'}-\phi_a) ]=0.405\nonumber\\
&&|\langle A_{1}B'_{2}|\Psi\rangle|^{2} = |\langle A_{2}B'_{1}|\Psi\rangle|^{2}={1\over 4}[1+\cos \theta_a\cos \theta_{b'} +\sin \theta_a\sin \theta_{b'} \cos(\phi_{b'}-\phi_a) ]=0.095.
\label{Born2}
\end{eqnarray}
For the eigenvectors of $(\boldsymbol{\sigma}\cdot {\bf a'}) \otimes (\boldsymbol{\sigma}\cdot {\bf b})$, we laso have: 
\begin{eqnarray}
&&|\langle A'_{1}B_{1}|\Psi\rangle|^{2}= |\langle A'_{2}B_{2}|\Psi\rangle|^{2}={1\over 4}[1-\cos \theta_{a'}\cos \theta_b -\sin \theta_{a'}\sin \theta_b \cos(\phi_b-\phi_{a'}) ]=0.095\nonumber\\
&&|\langle A'_{1}B_{2}|\Psi\rangle|^{2} = |\langle A'_{2}B_{1}|\Psi\rangle|^{2}={1\over 4}[1+\cos \theta_{a'}\cos \theta_b +\sin \theta_{a'}\sin \theta_b \cos(\phi_b-\phi_{a'}) ]=0.405.
\label{Born3}
\end{eqnarray}
Finally, for the eigenvectors of $(\boldsymbol{\sigma}\cdot {\bf a'}) \otimes (\boldsymbol{\sigma}\cdot {\bf b'})$, we have: 
\begin{eqnarray}
&&|\langle A'_{1}B'_{1}|\Psi\rangle|^{2}= |\langle A'_{2}B'_{2}|\Psi\rangle|^{2}={1\over 4}[1-\cos \theta_{a'}\cos \theta_{b'} -\sin \theta_{a'}\sin \theta_{b'} \cos(\phi_{b'}-\phi_{a'}) ]=0.095\nonumber\\
&&|\langle A'_{1}B'_{2}|\Psi\rangle|^{2} = |\langle A'_{2}B'_{1}|\Psi\rangle|^{2}={1\over 4}[1+\cos \theta_{a'}\cos \theta_{b'} +\sin \theta_{a'}\sin \theta_{b'} \cos(\phi_{b'}-\phi_{a'}) ]=0.405.
\label{Born4}
\end{eqnarray}

Clearly, we cannot have the four directions ${\bf a}$, ${\bf a'}$, ${\bf b}$ and ${\bf b'}$ to be coplanar, however, we can choose for example: $\phi_b = \phi_a=0$ and $\phi_{b'} =\phi_{a'}$. If we also set $\theta_a=0^\circ$, from (\ref{Born}) we obtain: ${1\over 4}(1-\cos \theta_b)= 0.095$, hence $\theta_b = \arccos 0.62 = 51.68^\circ$. In other words, {\bf b} makes an angle of $51.68^\circ$ with respect to ${\bf a}$. From (\ref{Born4}) we also have: ${1\over 4}[1- \cos (\theta_{b'} - \theta_{a'})]= 0.095$, hence $|\theta_{b'} - \theta_{a'}|=51.68^\circ$. We are thus considering a situation where the angles between ${\bf a}$ and ${\bf b}$, and between ${\bf a'}$ and ${\bf b'}$, are the same. From (\ref{Born2}) we have: ${1\over 4}(1-\cos \theta_{b'})= 0.405$, hence $\theta_{b'} = \arccos (-0.62) = 128.30^\circ$. Therefore, we also obtain: $\theta_{a'}=128.30^\circ - 51.68^\circ=76.62^\circ$. Finally, inserting these values in (\ref{Born3}), one finds (only writing approximate values for the angles):
%$0.231408\cdot 0.620053 +0.972857 \cdot 0.784560 \cos\phi_{a'} ]= 0.143472696 -+0.76326468 \cos\phi_{a'}=0.62$, so that $\cos\phi_{a'} = 0.62432744$
$\phi_{a'}=\phi_{b'}= 51.37^\circ$.

Replacing the above values in (\ref{eigenAB}), and in the corresponding eigenstates' expressions for the other product observables, we have thus obtained an explicit quantum representation of the data (\ref{data2}). Inserting these values also in (\ref{unitvectors}), again only writing approximate values for the different components, we have: 
\begin{eqnarray}
{\bf a}= \left[\begin{array}{c}
0\\ 0\\ 1
\end{array} \right],\quad 
{\bf b} =\left[\begin{array}{c}
0.78\\ 0\\ 0.62
\end{array} \right],\quad 
{\bf a'}= \left[\begin{array}{c}
0.61\\ 0.76\\ 0.23
\end{array} \right],\quad 
{\bf b'}= \left[\!\!\begin{array}{r}
0.49\\ 0.61\\ -0.62
\end{array} \right]. 
\label{spacedirections}
\end{eqnarray}
%a= (0, 0, 1); b=(0.78, 0, 0.62); a'= (0.61, 0.76, 0.23); b'= (0.49, 0.61, -0.62)
Let us also calculate the angles $\alpha_{{\bf a},{\bf b}}$, $\alpha_{{\bf a'},{\bf b}}$, $\alpha_{{\bf a},{\bf b'}}$ and $\alpha_{{\bf a'},{\bf b'}}$ between ${\bf a}$ and ${\bf b}$, ${\bf a'}$ and ${\bf b}$, ${\bf a}$ and ${\bf b'}$, and ${\bf a'}$ and ${\bf b'}$, respectively. We already observed that $\alpha_{{\bf a},{\bf b}}=\alpha_{{\bf a'},{\bf b'}} = 51.68^\circ$ (${\bf a}\cdot {\bf b}= {\bf a'}\cdot {\bf b'}=0.62$). Also: $\cos \alpha_{{\bf a},{\bf b'}} = {\bf a}\cdot {\bf b'} = -0.62$, so that $\alpha_{{\bf a},{\bf b'}} =128.30^\circ$. Finally, $\cos \alpha_{{\bf a'},{\bf b}} = {\bf a'}\cdot {\bf b} = 0.6184$, hence $\alpha_{{\bf a'},{\bf b}} =51.80^\circ$. Note also that $\alpha_{{\bf a},{\bf a'}}=76.70^\circ$ and $\alpha_{{\bf b},{\bf b'}}=90.13^\circ$.

\begin{figure}[htbp]
\begin{center}
\includegraphics[width=13cm]{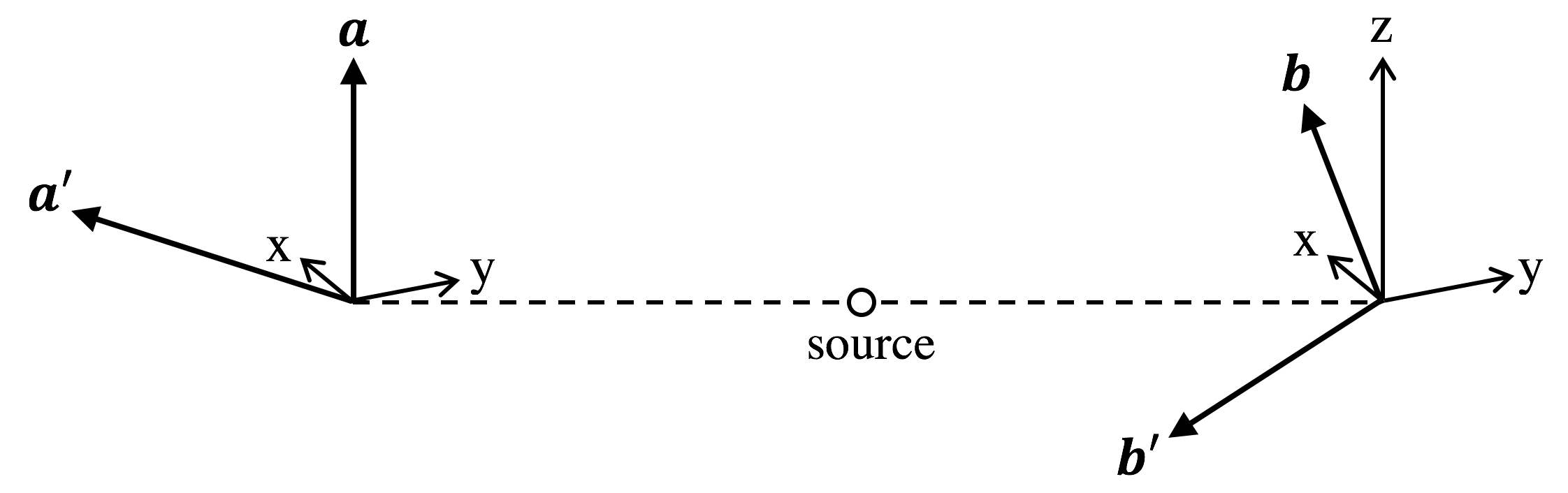}
\caption{The four unit vectors ${\bf a}$, ${\bf a'}$, ${\bf b}$ and ${\bf b'}$, describing the orientations of the Stern-Gerlach apparatuses that can reproduce the probabilities (\ref{data2}), when the four product measurements $(\boldsymbol{\sigma}\cdot {\bf a}) \otimes (\boldsymbol{\sigma}\cdot {\bf b})$, $(\boldsymbol{\sigma}\cdot {\bf a}) \otimes (\boldsymbol{\sigma}\cdot {\bf b'})$, $(\boldsymbol{\sigma}\cdot {\bf a'}) \otimes (\boldsymbol{\sigma}\cdot {\bf b})$ and $(\boldsymbol{\sigma}\cdot {\bf a'}) \otimes (\boldsymbol{\sigma}\cdot {\bf b'})$ are performed on bipartite entities prepared in the singlet state (\ref{singlet}).}
\label{Figure3}
\end{center}
\end{figure}

So, if we prepare a beam of composite (bipartite) spin-entities in the singlet state (\ref{singlet}), flying in opposite directions, and we perform the $A$-measurement ($A'$-measurement) on the left entity by orienting the Stern-Gerlach apparatus along the spatial directions ${\bf a}$ (${\bf a'}$), given by (\ref{spacedirections}), and similarly we perform the $B$-measurement ($B'$-measurement) on the right entity by orienting the Stern-Gerlach apparatus along the spatial directions ${\bf b}$ (${\bf b'}$), given by (\ref{spacedirections}), as illustrated in Fig.~\ref{Figure3}, then the statistics of outcomes of the joint measurements will exactly reproduce the probabilities (\ref{data2}). And this completes our quantum mechanical model, showing the great similarity between coincidence measurements in cognition, on combined concepts, and coincidence measurements on quantum entities (like spin entities) in entangled (singlet) states. 

\section{Response to criticisms\label{discussion}}

In this section, we provide some answers and commentaries to criticisms that have been raised about our approach in \cite{as2011, as2014} (see \cite{dk2014}) and in \cite{ass2017} (see \cite{dkczj2016}, where the authors refer to a previous preprint version of the present two-part article). More precisely, in \cite{dkczj2016} 
the authors maintain that, since it is always possible to transform a probability distribution violating the marginal law into an isotrophic one, in such a way that the transformation preserves the CHSH violation \cite{mag2016}, our analysis of Sec.~\ref{version} would not add much to the issue of the identification of genuine forms of entanglement in cognitive science. However, it should be observed that the symmetrization described in Sec.~\ref{version} is not a `data analysis procedure', meant to get rid of the marginal law in an {\it ad hoc} way, but the description of an experimental procedure defining  specific mixed measurements.

To obtain the data of these mixed measurements, we observed that they were already available,  under the hypothesis that participants would remain consistent in their choices. By `consistent' we mean that a person would choose in the same way in all measurements, i.e., if s/he chooses, say, {\it North} and {\it Northeast} in the unrotated $AB$ measurement, s/he will also choose {\it North} and {\it Northeast} in the corresponding $45^\circ$-rotated versions of the measurement, and so on. We are aware that this is an idealization; hence, we expect that in a real experiment there will be slight deviations from the probabilities we have calculated for the mixed experiment. However, they would be only due to the absence of a complete consistency of the participants' successive choices and are thus expected to be statistically insignificant. In other words, if the mixture of rotated joint measurements would be actually performed, it would typically preserve the marginal law. 

Anticipating a possible objection, one could say that a mixture of joint measurements is not a \emph{bona fide} joint measurement. Certainly it is not a standard joint measurement, but what is important is that experimentally speaking it is perfectly well-defined and that the outcomes are correlated in a significant way. In that respect, let us observe that alternative protocols are also possible. For instance, one could consider non-random mixtures: instead of rolling each time the die, one could decide to start with the unrotated measurements, then consider the $45^\circ$-rotated ones, then a further $45^\circ$-rotated version of them, and so on, proceeding in a cyclic way. 

What is important is to also understand the reason for defining such a mixed experiment, residing in the difference between the notions of `wind' and `space', as perceived by respondents in relation to the notion of `direction'. For example, the difference between {\it Northeast} and {\it Southeast}, which corresponds to an angle of $45^\circ$ from a pure spatial perspective, could be perceived as greater than for example the difference between {\it South East} and {\it Southwest}, which also corresponds to a spatial angle of $45^\circ$. Indeed, both {\it Southeast} and {\it Southwest} directions can be associated in Europe with a relatively warm weather, while between {\it Northeast} and {\it Southeast} the general perception is that of a more important change in terms of temperatures. In other words, participants, in their choice of pairs of different spatial directions, will be influenced by the latter being connected to winds, which is why the isotropy of space symmetry will be generally broken, which in turn provokes the violation of the marginal law. By defining an experiment formed by a uniform mixture of rotated measurements, the symmetry is restored whereas the violation of the CHSH inequality is preserved, showing that the former is not at all the consequence of the latter, being in structure completely independent of it. In that respect, we can hypothesize that if we had performed our experiment without connecting wind directions to space directions, hence directly considering pure space directions, we would have equally observed a violation of the CHSH inequality, without however introducing the asymmetry that is at the origin of the violation of the marginal law.

In order to make our point even more explicit, we will consider in Sec.~\ref{appendix} the example of a macro-physical composite system violating both the CHSH inequality and the marginal law, with the latter being clearly the consequence of a lack of symmetry/isotropy in the experiment, so much so that when the full symmetry of the apparatuses is restored, the marginal law is recovered whereas the violation of the CHSH inequality is maintained, thus showing that the latter is the result of a mechanism (the possibility for the sub-systems to remain connected and influence each other) that is independent of a possible lack of symmetry producing a violation of the marginal law.  In fact, in the ambit of this model the latter tends to reduce instead of increase the magnitude of the violation.\footnote{The example given in Sec.~\ref{appendix} is a generalization of the mechanistic classical laboratory situation presented in \cite{Aerts1991, aabg2000}, which was more recently also analyzed in \cite{AertsSassoli2016}, in the ambit of so-called extended Bloch representation of quantum theory.} 

Another criticism expressed in \cite{dkczj2016} is that if we define entanglement as any violation of the CHSH inequality, this would make the construction of entangled systems a mere ``child's play,'' in the sense that they would become ``ubiquitous and obvious,'' to use the authors' words. This is however exactly the point: entangled systems are ubiquitous in micro-physics because of the superposition principle, according to which each time we have two product states, we can superpose them and generate an infinity of entangled states. In that respect, it is worth mentioning that it is precisely because of this that the standard quantum formalism is unable to describe experimentally separated entities, as demonstrated by one of us in the eighties of last century \cite{a1984}. Would entanglement then become obvious? Well, this is certainly so for those systems where the nature of the connection allowing the systems to communicate presents no mystery. This is the case of all macroscopic systems whose violation of the CHSH inequality is due to the presence of a spatial connection that allows the sub-systems to work as a whole (and in that sense to influence each other; see the paradigmatic example presented in Sec.~\ref{appendix}, where the connection between the two sub-systems is made manifest by the presence of a third `element of reality': the rigid rod). 

For micro-physical quantum entangled systems the situation is less obvious, because the `connective element of reality' responsible for the creation of correlations cannot be directly detected in our spatiotemporal theater. This because quantum entities are non-spatial entities, only acquiring spatial properties when submitted to specific measurement processes. However, if one adopts the recently developed extended Bloch representation of quantum theory \cite{AertsSassoli2016}, it is possible to generally and consistently represent the state of a bipartite entity as a triple of real vectors $(r_1,r_2,r_{12})$, where $r_1$ and $r_2$ are the Bloch vectors specifying the states of the two sub-systems, and the third (higher dimensional) vector $r_{12}$  describes the  state of their connection. This means that the general idea that entanglement would be a process of `creation of correlations' due to the presence of a connective element between the sub-systems, remains valid also for micro-physical entities, although in their case the latter cannot be understood as a mere `connection through space'.

It is worth emphasizing that the identification of a `connective element of reality' associated with entanglement in the extended Bloch representation should not be considered as a mere artefact of mathematical nature. For a spin-${1 \over 2}$ quantum entity (qubit), the Bloch representation is three-dimensional and allows to associate a space direction to each spin state. Therefore, when used to also describe measurements (i.e., the interaction with Stern-Gerlach apparatuses), one obtains a representation that is as close as possible to a spatial representation, thanks to the connection between $SU(2)$ and the rotation group $SO(3)$. When considering higher-dimensional Hilbert spaces, like the 4-dimensional Hilbert space of a bipartite system formed by two spin-${1 \over 2}$ entities, the extended Bloch representation still allows to describe the individual spin states as directions in the 3-dimensional Euclidean space. So, it is the representation one needs to adopt if willing to have a description of the measurement processes as mechanisms taking place `as close to space as possible'. Then, following our analysis in \cite{AertsSassoli2016}, one can show that for the representation to remain consistent with the observed correlations (as predicted by the Born rule), the connection between the two sub-entities generally requires more than the three spatial dimensions to be properly described, hence it is a non-spatial connection (the situation described in Sec.~\ref{appendix} is in that sense an exception, only valid when the initial state is prepared in a rotationally invariant singlet state). 

What about human conceptual entities? The situation is certainly more obvious in their regard for the following reason: human minds are perfectly capable to handle abstractions and detect when there is a `meaning connection' between two concepts. When this connection is strong enough, significant correlations can be created, which in turn can violate the CHSH inequality. If we consider the general `quantum cognition hypothesis', i.e., that human concepts would exhibit, in their combinations and interactions with human minds, a quantum-like organization, then as well as entanglement is ubiquitous among micro-entities, likewise it must also be ubiquitous in phenomena such as information processing, decision making, concepts and conceptual reasoning, human judgment, etc. 

To further clarify our motivation in finding deep similarities between the entanglement we identified in cognitive experiments on combinations of concepts \cite{ass2017,as2011,as2014} and the entanglement identified in physics laboratories, we can also mention our ongoing investigation about an explanatory framework for quantum theory that we called the `conceptuality interpretation' \cite{Aerts2009b, Aerts2010a, Aerts2010b, Aerts2013, Aerts2014}. The main hypothesis put forward in this new interpretation, with far-reaching consequences for many aspects of our physical reality, is that micro-physical entities would be also conceptual-like, in the sense of behaving more like conceptual (meaning) entities than like objects. However, the conceptuality/quantumness of microphysical entities would also be different from the conceptuality/quantumness of human concepts. Indeed, quoting from \cite{Aerts2010a}: 

``[$\cdots$] human concepts and their interactions are at a very primitive stage of development as compared to quantum entities and their interactions. This means that, although we expect to find connections with a profound explanatory potential with respect to fundamental aspects of both situations, i.e. human concepts and their interactions and quantum particles and their interactions, we also expect to find a much less crystallized and organized form for human concepts than for quantum particles. [$\cdots$] This means that we regard the actual structure of the physical universe, space, time, momentum, energy and quantum particles interacting with ordinary matter as emergent from a much more primitive situation of interacting conceptual entities and their memories. Consequently, [$\cdots$] one of the research aims must be to investigate which structural properties, laws and axioms may characterize a weakly organized conceptual structure, such as the one actually existing for the case of human concepts and memories, and which additional structural properties, laws and axioms could make it into a much more strongly organized conceptual structure, such as the one of the physical universe, space time, momentum energy and quantum particles interacting with ordinary matter.''

The violation of the marginal law, which is typical in experimental situations with human concepts interacting with human minds, is precisely an example of this weaker (in the sense of exhibiting lesser symmetry) organization, in comparison to micro-physical entities interacting with measuring apparatuses, like the Stern-Gerlach ones in spin measurements. However, a weaker organization/symmetry does not necessarily mean a weaker connection between the concepts forming a composite conceptual entity, and if entanglement is understood as the presence of such connection, then the generalization of the CHSH inequality considered in \cite{dk2014,dkczj2016}, where deviations from classicality due to violations of the marginal law are subtracted from the usual CHSH factor, would not 
have general validity in order to 
ascertain the presence of entanglement.

Of course, different definitions of entanglement (or contextuality, to use the terminology in \cite{dk2014, dkczj2016}) are possible, so we are not affirming here that the modified CHSH inequality used in \cite{dk2014, dkczj2016} would not be of interest, both for physicists and psychologists. What we are saying is that it can hardly capture all facets of the entanglement phenomenon, as our symmetrized version of the \emph{Two Different Wind Directions} experiment and the mechanical example described in Sec.~\ref{appendix} clearly show. In other words, the notion of contextuality is defined in \cite{dk2014,dkczj2016}  having in mind that not all ways to connect or to communicate should be part of the definition of entanglement. For instance, systematic human errors could produce a violation of the marginal law in physics, and one may indeed want to exclude (filter out) these `measurement biases due to systematic errors' from the calculation of a degree of contextuality/entanglement. This is precisely what the modified CHSH inequality in \cite{dk2014,dkczj2016} does. But one should also take care not to throw the baby away with the dirty water. Indeed, the marginal law can be violated for 
%M very 
many 
different reasons, and 
%M 
systematically 
subtracting the magnitude of its violation from the CHSH value might not always be the right thing to do.

\section{The generalized rigid rod model\label{appendix}}

In this section, we present a model consisting of two generalized entangled (interconnected) `quantum machines'. This is a composite system generating a probability model that can go beyond that of standard quantum mechanics. Its interest, among other things, lies in the fact that it allows to study the marginal law in a quantum-like context where it can be easily violated. In this way, one can show that not only the marginal law can be obeyed by non-Hilbertian models, but also that its violation can result from a lack of symmetry of the measurements that are operated on the sub-systems. An additional element of interest of the model is that the `connective element of reality' that is responsible for the violation of the CHSH inequality is clearly identifiable in it, and also perfectly distinguishable from what produces the violation of the marginal law, in accordance with our analysis in the previous sections.

Each `quantum machine' is an empty unit sphere containing a point particle, describing the quantum state (according to the Bloch sphere representation). Measurements are performed by means of elastic bands, stretched along the diameters of the sphere. More specifically, the point particle first reaches the elastic, following a path orthogonal to it, then remains attached to it, and when the elastic breaks (in some unpredictable point), the particle is drawn towards one of its two end points, corresponding to the outcome (state) of the measurement. If the elastic is uniform, the outcome probabilities are exactly those predicted by the Born rule. If it is not uniform, different (more general) `probabilistic rules' can be described (for more details, see for instance \cite{AertsSassoli2014c, AertsSassoli2015}). 

\begin{figure}[!ht] \centering \includegraphics[scale =.5]{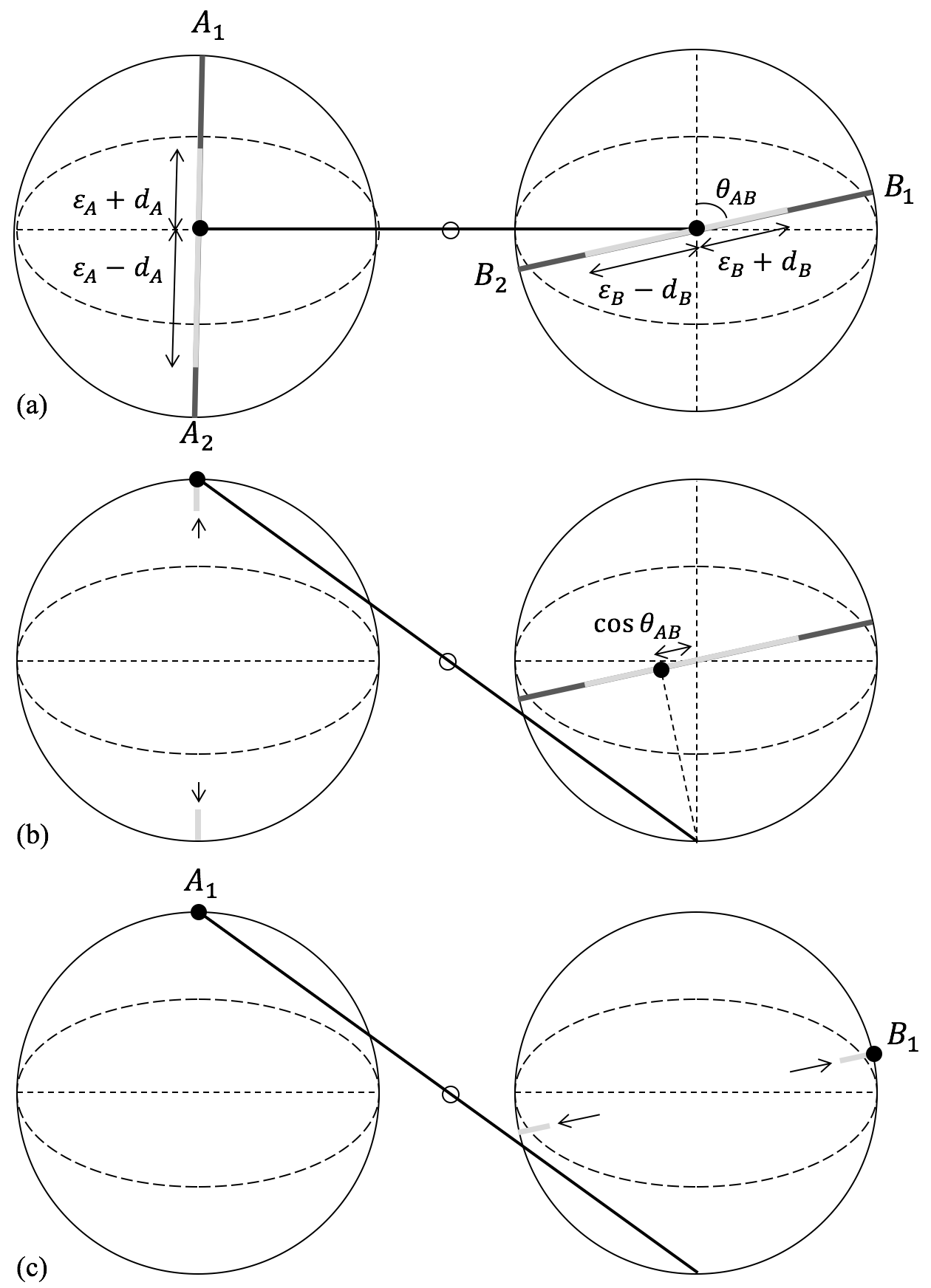} 
\caption{(a) Initially the two particles are at the center of their respective spheres; (b) following the $A$-measurement, the left particle is drawn either to $A_1$ or to $A_2$, with equal probability; here the outcome is $A_1$; because of the rod-connection, the right particle is forced to acquire the opposite position in its own sphere. The rod-connection is then disabled and the right particle orthogonally ``falls'' onto the elastic band associated with the $B$-measurement; (c) the right particle is finally drawn either to $B_1$ or $B_2$ (here $B_1$). Note that we are here in the situation where $d_A$ and $d_B$ are both taken to be negative. \label{rod}} 
\end{figure}

Two quantum machines can be entangled by connecting the point particles (assumed to be both initially at the center of their respective spheres) by means of a rigid rod of variable length, rotating around a pivot, always keeping the two particles at the same distance from the latter (see Figure~\ref{rod}). Because of the presence of this connecting element, measurement outcomes will exhibit correlations, able to violate the CHSH inequality. In particular, if all elastics are uniform (i.e., they all break with same probability in all points), this bipartite mechanistic entity behaves exactly like two spin-${1\over 2}$ entangled entities in a singlet state, and thus violates (for a suitable choice of the angles between the elastics) the CHSH inequality with the exact $2\sqrt{2}$ maximal quantum value \cite{Aerts1991, aabg2000,AertsSassoli2016}). 

We want to analyze what are the joint probabilities associated with this rigid rod model in the general situation where the four measurements considered are characterized by non-uniform elastic bands. More precisely, we consider four measurements $A$, $A'$, $B$ and $B'$, defined by locally uniform elastic bands described by the parameters $(\epsilon_A,d_A)$, $(\epsilon'_A,d'_A)$, $(\epsilon_B,d_B)$ and $(\epsilon'_B,d'_B)$, and relative angles $\theta_{AB}$, $\theta_{A'B}$, $\theta_{AB'}$ and $\theta_{A'B'}$, respectively. The meaning of the two parameters $\epsilon_A \in [0,1]$ and $d_A\in [-1+\epsilon_A, 1-\epsilon_A]$ is the following: an $(\epsilon_A,d_A)$-elastic is such that it is uniformly breakable along its internal segment $[d_A-\epsilon_A, d_A+\epsilon_A]$, and unbreakable everywhere else, with the values $1$ and $-1$ corresponding to the two elastic's end points $A_1$ and $A_2$, respectively, and same for the other measurements. The $A$ and $A'$ measurements are performed on the point particle in the left sphere, whereas the $B$ and $B'$ measurements are performed on the point particle in the right sphere. To simplify the discussion, we also assume that the above four angles are such that the point particle in the second measurement always falls onto a breakable region of the elastics. 

We start by calculating the probabilities associated with the sequential measurement `$A$ then $B$'. Consider the outcome $(A_1,B_1)$. First the left elastic breaks, drawing the point particle towards the end point $A_1$ with probability ${\epsilon_A - d_A\over 2\epsilon_A}$, given by the ratio between the length of the elastic segment whose breaking causes the point particle to reach $A_1$, which is $\epsilon_A - d_A$, and the total length of the (uniformly) breakable segment of the elastic, which is $2\epsilon_A$. When the point particle reaches the $A_1$ end point, because of the rod-connection the right particle is forced to acquire the opposite position in its own sphere. At this point the rod-connection is disabled, and the right particle orthogonally ``falls'' onto the elastic band associated with the $B$-measurement, reaching a position that depends on the relative angle $\theta_{AB}$ between the two elastic bands (see Figure.~\ref{rod}). The $(\epsilon_B,d_B)$-elastic then also breaks in an unpredictable point, drawing the right point particle either towards $B_1$ or $B_2$, the probability for the former being given again by the ratio between the length of the elastic segment whose breaking causes the point particle to reach $B_1$, which is $\epsilon_B - d_B -\cos\theta_{AB}$, and the total length of the (uniformly) breakable segment of the elastic, which is $2\epsilon_B$. Reasoning in the same way for the other three pairs of outcomes, we thus obtain the formulae: 
\begin{eqnarray}
&p_{A\to B}(A_1,B_1) =\left({\epsilon_A-d_A\over 2\epsilon_A}\right)\left({\epsilon_B-d_B-\cos\theta_{AB}\over 2\epsilon_B}\right),\quad 
p_{A\to B}(A_1,B_2) = \left({\epsilon_A-d_A\over 2\epsilon_A}\right)\left({\epsilon_B+d_B+\cos\theta_{AB}\over 2\epsilon_B}\right)\nonumber\\
&p_{A\to B}(A_2,B_1) = \left({\epsilon_A+d_A\over 2\epsilon_A}\right)\left({\epsilon_B-d_B+\cos\theta_{AB}\over 2\epsilon_B}\right),\quad 
p_{A\to B}(A_2,B_2) = \left({\epsilon_A+d_A\over 2\epsilon_A}\right)\left({\epsilon_B+d_B-\cos\theta_{AB}\over 2\epsilon_B}\right).
\label{AB-prob}
\end{eqnarray} 
For the reversed order sequential measurement `$B$ then $A$', we can calculate in the same way the joint probabilities and we obtain: 
\begin{eqnarray}
&p_{A\leftarrow B}(A_1,B_1) = \left({\epsilon_B-d_B\over 2\epsilon_B}\right)\left({\epsilon_A-d_A-\cos\theta_{AB}\over 2\epsilon_A}\right),\quad 
p_{A\leftarrow B} (A_1,B_2)= \left({\epsilon_B+d_B\over 2\epsilon_B}\right)\left({\epsilon_A-d_A+\cos\theta_{AB}\over 2\epsilon_A}\right)\nonumber\\
&p_{A\leftarrow B}(A_2,B_1) = \left({\epsilon_B-d_B\over 2\epsilon_B}\right)\left({\epsilon_A+d_A+\cos\theta_{AB}\over 2\epsilon_A}\right),\quad
p_{A\leftarrow B}(A_2,B_2) = \left({\epsilon_B+d_B\over 2\epsilon_B}\right)\left({\epsilon_A+d_A-\cos\theta_{AB}\over 2\epsilon_A}\right).
\label{BA-prob}
\end{eqnarray} 

Assuming that we don't know in which order the measurements are performed, we can consider a uniform mixture of the above probabilities: $
{\bar p}(A_i,B_j)={1\over 2} [p_{A\to B}(A_i,B_j) +p_{A\leftarrow B}(A_i,B_j)]$, 
so that we obtain:
\begin{eqnarray}
&{\bar p}(A_1,B_1) = {1\over 8}(1-{d_A\over\epsilon_A})(1-{d_B+\cos\theta_{AB}\over\epsilon_B})+{1\over 8}(1-{d_B\over\epsilon_B})(1-{d_A+\cos\theta_{AB}\over\epsilon_A})\nonumber\\
&{\bar p}(A_1,B_2) = {1\over 8}(1-{d_A\over\epsilon_A})(1+{d_B+\cos\theta_{AB}\over\epsilon_B})+{1\over 8}(1+{d_B\over\epsilon_B})(1-{d_A -\cos\theta_{AB}\over\epsilon_A})\nonumber\\
&{\bar p}(A_2,B_1) = {1\over 8}(1+{d_A\over\epsilon_A})(1-{d_B -\cos\theta_{AB}\over\epsilon_B})+{1\over 8}(1-{d_B\over\epsilon_B})(1+{d_A +\cos\theta_{AB}\over\epsilon_A})\nonumber\\
&{\bar p}(A_2,B_2) = {1\over 8}(1+{d_A\over\epsilon_A})(1+{d_B -\cos\theta_{AB}\over\epsilon_B})+{1\over 8}(1+{d_B\over\epsilon_B})(1+{d_A -\cos\theta_{AB}\over\epsilon_A}),
\end{eqnarray} 
and similar expressions hold for the other measurements. It follows that the marginal probabilities are: 
\begin{eqnarray}
&{\bar p}(A_1,B_1)+{\bar p}(A_1,B_2)={\epsilon_A -d_A\over 2\epsilon_A} +{d_B\cos\theta_{AB}\over 4\epsilon_A\epsilon_B},\quad 
{\bar p}(A_2,B_1)+{\bar p}(A_2,B_2)={\epsilon_A +d_A\over 2\epsilon_A} -{d_B\cos\theta_{AB}\over 4\epsilon_A\epsilon_B}\nonumber\\
&{\bar p}(A_1,B_1)+{\bar p}(A_2,B_1)={\epsilon_B - d_B \over 2\epsilon_B}+ {d_A\cos\theta_{AB}\over 4\epsilon_A \epsilon_B},\quad {\bar p}(A_1,B_2)+{\bar p}(A_2,B_2)={\epsilon_B + d_B \over 2\epsilon_B}- {d_A\cos\theta_{AB}\over 4\epsilon_A \epsilon_B}.
\end{eqnarray} 

The first terms in the above four expressions are what we expect to obtain for the marginal probabilities in case the two particles would be experimentally separated, that is, in case the rod mechanism would not be present. The second terms are the contributions responsible for the violation of the marginal law. Note that they are non-zero even if the two elastics are the same, i.e., $(\epsilon_A,d_A)=(\epsilon_B,d_B) =(\epsilon,d)$, which is the situation we are now going to consider, for simplicity. This is so because if they are non-symmetric with respect to the origin of the spheres ($d\neq 0$), considering that the rod produces anticorrelations instead of correlations, order effects will still be present. 

Following a simple calculation, one obtains for the expectation value: 
\begin{equation}
E(A,B)=-{\cos\theta_{AB}\over\epsilon} + {d^2\over\epsilon^2},
\end{equation}
and similar formula also hold for $E(A,B')$, $E(A',B)$ and $E(A',B')$. Note that in the limit of uniformly breakable elastic bands $(\epsilon,d)\to (1,0)$, one recovers the pure quantum expectation value: $E(A,B)=-\cos\theta_{AB}$. Considering then the situation where the elastics are all coplanar and have relative angles $\cos\theta_{AB}= \cos\theta_{AB'}=\cos\theta_{A'B'}=-\cos\theta_{A'B}={\sqrt{2}\over 2}$ (this is the situation producing the maximum violation of the CHSH inequality in quantum theory), we find the expectation values: 
\begin{equation}
E(A,B)=E(A',B)=E(A',B')=-{\sqrt{2}\over 2\epsilon} + {d^2\over\epsilon^2},\quad E(A,B')={\sqrt{2}\over 2\epsilon} + {d^2\over\epsilon^2}, 
\end{equation}
so that $|S| = |E(A,B)-E(A,B') + E(A',B') + E(A',B)|$ becomes:
\begin{equation}
|S| =2\left|{\sqrt{2}\over\epsilon}- {d^2\over\epsilon^2}\right|.
\label{CHSH-modified}
\end{equation}

In the special case where the elastics are symmetric ($d=0$), but not necessarily globally uniform ($\epsilon \leq 1$), we have $|S| = {2\sqrt{2}\over \epsilon}$, and of course the marginal law is preserved. As we assumed that the particle always land onto a breakable fragment, we have the condition $|\cos\theta_{AB}|\leq \epsilon$ (and same for the other angles), so that $\epsilon \geq {\sqrt{2}\over 2}$. Inserting the minimal value $\epsilon={\sqrt{2}\over 2}$ in the previous formula, we thus obtain $|S| =4$. In other words, it is possible to maximally violate the CHSH inequality and at the same time preserve the marginal law. In fact, as is clear from (\ref{CHSH-modified}), the violation of the latter, which results from the asymmetry of the elastics, goes in the direction of reducing, instead of increasing, the magnitude of the violation.

\section{Conclusion\label{conclusions}}

In this (second half of our two-part) article we have considered a refined version of the experiment presented in \cite{ass2017}, which violates (the CHSH version of) Bell's inequality and at the same time preserves the marginal law, thus showing that human minds select wind directions in ways that are similar to how Stern-Gerlach apparatuses (polarizers) select spin directions (photon's directions of polarization) on bipartite spin (photonic) entities in singlet states. 

More precisely, by analyzing the lack of symmetry of our previous experimental situation, we have formulated the hypothesis that the violation of the marginal law is in the present situation just the consequence of such symmetry breaking, so that the violation of the marginal law should not be considered as an argument against the presence of a genuine form of entanglement between the two conceptual sub-systems, to be interpreted as the presence of an abstract meaning connection between the latter. Furthermore, we have presented a full quantum modeling in Hilbert space of the experimental data and have also supported our conclusions by presenting a model consisting of two asymmetric interconnected `quantum machines', where the independent role played by the marginal law and the CHSH inequality is explicitly revealed. 

Our analysis, by providing an example of entanglement in cognition that is much closer to the one encountered in physics, also provides an alternative explanation of the observed marginal law violations in physics laboratories, which is that the assumption that the Stern-Gerlach apparatuses, or the polarizers, perfectly obey the isotropy of space symmetry would not necessarily always be valid in real experimental situations. If this hypothesis turns out to be correct, which of course will require further investigations, it would make a good example of how a comparison between the way data are collected in cognition and physics can possibly shed light on both experimental situations. 

At the present stage of our understanding of the micro-physical realm, it is not clear if the observed violations of the marginal law should be attributed to experimental defects or would instead reveal a more fundamental `lack of spatial symmetry' of the measurement context. In the second case, it is also not clear if the observed violations would be due to the fact that the joint measurements are not be reducible to separate measurements on the individual components, or that the individual measurements exhibit some kind of order effects. As regards the first hypothesis, note that non-product measurements on composite systems are operationally well defined in quantum physics, and mathematically described in terms of `entangled (non-separable) measurements'. They have been widely explored in quantum cryptography and teleportation, and are useful resources in quantum information. In fact, we precisely used a combination of entangled states and entangled measurements to faithfully represent the experimental data of the {\it The Animal Acts} experiment \cite{as2014}, and the same we could have done for the data (\ref{data}) of the unsymmetrized experiment \cite{ass2017}.  In other words, when the  marginal law is violated, it is possible to describe measurements on bipartite conceptual entities as entangled measurements, and this alternative methodological approach has its own independent justification. 

Therefore, the violation of the marginal law, instead of being interpreted as something going in the direction of a reduction of entanglement, according to the specific  definition of contextuality proposed in \cite{dk2014, dkczj2016}, can also be associated, contrariwise, with the presence of a stronger form of it, not only associated with states, but also with measurements. In other words, even in situations where the marginal law is violated, experiments like ours can be interpreted as revealing a genuine form of entanglement in behavioral data. This however requires a definition of entanglement that, unlike the one adopted in \cite{dk2014, dkczj2016}, does not {\it a priori} exclude its interpretation as a `form of connection between the parts of a composite system'. In our view, this is what entanglement precisely is: a non-ordinary connection emerging when micro-physical systems are combined, which cannot be described as a `connection through space', whose counterpart in conceptual combinations is the `meaning connection' that systematically link the different concepts. 

What about the possibility of explaining the marginal law violations in terms of order effects? Order effects can only be modeled if the measurements performed by Alice and Bob are described as incompatible measurements. However, to model the experimental data it will be generally insufficient to use standard Hermitian operators. Instead, one has to go beyond-quantum and introduce generalized measurements, like those describable in the ambit of the `general tension-reduction (GTR) model', as recently emphasized in \cite{asdb2017,abssv2017}. In that respect, it would be  interesting to design ``sequentialized versions'' of the {\it The Animal Acts} and {\it Two Different Wind Directions} experiments, to see  if the violation of the marginal law that is obtained in this way will be of the same magnitude of the one produced by genuine coincidence measurements.

A last remark is in order. One may wonder why we have used, as the equivalent of the pair of Stern-Gerlach apparatuses in a laboratory EPRB experiment (or pair of polarizers, if photons are considered), the mind of a single person, and not a pair of human minds, associated with two distinct persons. Clearly, if we would have performed our experiment using a two-mind experimental context, with each mind only selecting a wind direction, also making sure that they have no way whatsoever to communicate during the process, then the CHSH inequality would not be violated, and consequently entanglement could not be identified. This is so because for two separated minds (two persons that cannot communicate and therefore consult each other about their possible choices) the selection of a single wind direction is a rather meaningless process if it is meant to represent their best example of two different wind directions. Indeed, for the process to be truly meaningful, both minds would need to know about each other choices. Alternatively, one could ask the two minds to both select a pair of different wind directions, but only consider one direction for each of them, to be then combined in order to form the final outcome, but also in this case the CHSH inequality would not be violated. 

The problem is that when we consider two separate minds, we automatically also consider two {\it  Two Different Wind Directions} conceptual entities, one for each of them. The analogy with physics is then the situation where each Stern-Gerlach apparatus would perform the spin measurement on a distinct bipartite spin entity in a singlet state, and of course also in this situation the CHSH inequality would not be violated. Here it is important to distinguish the printable words that we use to denote a concept, and the concept itself, understood as a meaning entity belonging to an abstract mental realm. It is the {\it Two Different Wind Directions} conceptual entity that interacts with the human mind, not the spatiotemporal physical signs that we use to indicate it, i.e., the English words `Two Different Wind Directions' that we can write on a piece of paper, which are only a collapsed version of the former. Of course, a human mind will first read (through its eyes and brain) the written words, and by doing so it will ``cloth" them with meaning, i.e., associate them with an abstract {\it Two Different Wind Directions} conceptual entity, which we denote using the Italic font precisely to avoid confusing it with the former.\footnote{Concerning the important distinction between printable words/texts and the conceptual (meaning) entity associated with them, see \cite{QWeb2017}.}

Now, two distinct and separated mind-brain systems, interacting with the `Two Different Wind Directions' combination words, will create two distinct {\it Two Different Wind Directions} conceptual entities, not a single one. Hence, if we use two minds instead of one, we are in the situation where there are two composite conceptual entities submitted to measurements, not a single one. Thus, if we want an experiment in cognition to be as close as possible to a typical EPRB-experiment in physics, it is the `one-mind situation' and not the `two-mind situation' that we need to consider. In other words, the correspondence `one mind--two Stern-Gerlach apparatuses' is more appropriate than the correspondence `two minds--two Stern-Gerlach apparatuses'. Does this mean that two Stern-Gerlach apparatuses, although spatially separated, form one whole physical system, when interacting with a composite entity in an entangled state? We believe that indeed two Stern-Gerlach apparatuses can give rise to a `single domain of coherence' (where the notion of `coherence' plays the same role in physics as the notion of `meaning' in human cognition), whenever they interact with bipartite entities in entangled states, and that this is precisely at the origin of the violation of the CHSH inequality.

Our example of the two connected `quantum machines' can again be of help in explaining why physical apparatuses can easily form such `one domain of coherence', when they interact with two physical entities which, in turn, are interconnected by means of a third element, such as the rigid rod. Indeed, physical connections easily transport coherence between possibly distant systems, which otherwise could not behave as a coherent whole. For human minds interacting with concepts, via the input of written words, this same coherence/meaning will be naturally present within the semantic space generated by each individual mind, and when more than a single mind is involved, coherence/meaning can only appear if a third element is also allowed to play a role, which is the element of the `between minds conversation'.

And indeed, if two minds can communicate and form a unified semantic realm, an experiment using two minds will be able to also violate the CHSH inequality, by means of the meaning connection between the two conceptual elements {\it One Wind Direction} and {\it Another Wind Direction}. Of course, it is necessary for this that within this unified/interconnected semantic realm the communication is such that it can reveal which one of the two wind directions has been chosen by both minds, so that they can co-decide about which pair of directions should be selected as the best example of {\it Two Different Wind Directions}. We know that in traditional Bell-test experiments the question of determining whether Alice and Bob can communicate is an important and crucial aspect, but we should not confuse this with the issue we are considering here. Indeed, in micro-physics, the possible communication between Alice and Bob is not what is responsible for the violation (we can easily see this by considering again the rigid rod example), while in a two-mind cognitive experiment it is precisely what would be at the origin of it.

\end{document}